\documentclass[aip,graphicx,amsmath,preprint,unsortedaddress,eqsecnum,showpacs]{revtex4}

\usepackage{graphicx}
\usepackage{dcolumn}
\usepackage{bm}
\usepackage{float}
\usepackage{color}
\usepackage{amsfonts}

\begin{document}

\title{Roaming dynamics in ion-molecule reactions: phase space reaction pathways 
and geometrical interpretation}

\author{Fr\'ed\'eric A. L. Maugui\`{e}re}
\email{frederic.mauguiere@bristol.ac.uk}
\affiliation{School of Mathematics, University of Bristol, Bristol BS8 1TW, United Kingdom}

\author{Peter Collins}
\email{peter.collins@bristol.ac.uk}
\affiliation{School of Mathematics, University of Bristol, Bristol BS8 1TW, United Kingdom}

\author{Gregory S. Ezra}
\email{gse1@cornell.edu}
\affiliation{Department of Chemistry and Chemical Biology, Baker Laboratory, 
Cornell University, Ithaca, NY 14853, United States}

\author{Stavros C. Farantos}
\email{farantos@iesl.forth.gr}
\affiliation{Institute of Electronic Structure and Laser, Foundation for Research and Technology - Hellas, and \\
Department of Chemistry, University of Crete, Iraklion 711 10, Crete, Greece }

\author{Stephen Wiggins}
\email{stephen.wiggins@mac.com}
\affiliation{School of Mathematics, University of Bristol, Bristol BS8 1TW, United Kingdom}
\date{\today}

\begin{abstract}
A model Hamiltonian for the reaction 
CH$_4^+ \rightarrow$  CH$_3^+$ + H, parametrized to exhibit either early or late 
inner transition states, is employed to investigate the dynamical characteristics of the roaming 
mechanism.
Tight/loose transition states and conventional/roaming reaction pathways are 
identified in terms of time-invariant objects in phase space. 
These are dividing surfaces associated with normally hyperbolic invariant manifolds (NHIMs).
For systems with two degrees of freedom 
NHIMS are unstable periodic orbits which, in conjunction with their stable and unstable
manifolds, unambiguously define the (locally) 
non-recrossing dividing surfaces assumed in statistical theories of reaction rates.
By constructing periodic orbit continuation/bifurcation diagrams for two 
values of the potential function parameter  
corresponding to late and early transition states, respectively, and using the total energy as another parameter, 
we dynamically assign different regions of phase space to reactants and products as well as to
conventional and roaming reaction pathways. 
The classical dynamics of the system are investigated by uniformly sampling 
trajectory initial conditions on the dividing surfaces.
Trajectories are classified into four different categories: direct reactive and non reactive trajectories,
which lead to the formation of molecular and radical products respectively, and roaming reactive and non reactive 
orbiting trajectories, which represent alternative pathways to form molecular and radical products.
By analysing gap time distributions at several energies we demonstrate that 
the phase space structure of the roaming region,
which is strongly influenced by non-linear resonances between the two degrees of freedom, 
results in nonexponential (nonstatistical) decay. 
\end{abstract}


\maketitle 


\section{Introduction}
\label{sec:intro}

New experimental techniques for studying chemical reaction dynamics, such as imaging methods \cite{Ashfold06} and 
multidimensional infra-red spectroscopy \cite{Mukamel2000}, 
have revealed unprecedented details of the mechanisms of chemical reactions. 
The temporal and spatial resolution achieved 
allows the measurement of reactant and product state distributions, 
thus providing data that challenge existing theory. Given that accurate quantum dynamical 
studies can be carried out only for small polyatomic molecules, most theoretical analyses of chemical 
reaction rates and mechanisms are 
formulated in terms of classical mechanics (trajectory studies) 
or  statistical approaches, such as RRKM (Rice, Ramsperger, Kassel and Marcus) theory \cite{Forst03,Baer96}
or transition state theory (TST) \cite{Levine09}.

A significant challenge to conventional 
approaches to reaction mechanism is provided by the recently discovered ``{\it roaming reactions}''. 
This type of reaction was revealed in 2004 by Townsend \textit{et al.} 
in a study of photodissociation of formaldehyde \cite{townsend2004roaming}.
When excited by photons, the formaldehyde molecule can dissociate via two channels: 
H$_2$CO $\rightarrow$ H + HCO 
(radical channel) or H$_2$CO $\rightarrow$ H$_2$ + CO (molecular channel). 
Zee \textit{et al.} \cite{zee:1664} found
that, above the threshold for the H + HCO dissociation channel, 
the CO rotational state distribution exhibited an intriguing 
`shoulder' at lower rotational levels correlated with a hot 
vibrational distribution of H$_2$ co-product. The observed product 
state distribution did not fit well with the traditional picture of the dissociation of 
formaldehyde via the well characterized 
saddle point transition state for the molecular channel. 
Instead, a new pathway is followed that is dynamical in nature, 
and such dynamical reaction paths or roaming mechanisms
are the central topic of this paper.

The roaming mechanism, which explains the observations of Zee 
and co-workers, was demonstrated both experimentally and in classical trajectory
simulations by Townsend   \textit{et al.} 
\cite{townsend2004roaming}. Following this work, roaming  
has been identified in the unimolecular dissociation of molecules 
such as CH$_3$CHO, CH$_3$OOH or CH$_3$CCH,  and in  ion-molecule reactions \cite{Yu11}, and is now recognized as a 
general phenomenon in unimolecular decomposition (see Ref. [\onlinecite{bowman2011roaming}] and references therein).

Reactions exhibiting roaming pose a considerable 
challenge to basic understanding concerning the dynamics of 
molecular reactions. 
The standard picture in reaction dynamics is firmly based on  the concept of the reaction coordinate
\cite{Heidrich95}, for example, the {\it intrinsic reaction coordinate} (IRC). 
The IRC is a minimum energy path (MEP) in 
configuration space that smoothly connects reactants to products 
and, according to conventional wisdom, it is the path
a system follows (possibly modified by small fluctuations about this path) as reaction occurs. 
Roaming reactions, instead, avoid the IRC and involve 
alternative reaction pathways. 
(It is important to note that reactions involving dynamics
that avoids the IRC, so-called non-MEP reactions, 
were extensively studied before the term ``roaming'' was coined \cite{Sun02,Lopez07,Mikosch08,Zhang10}.)

For the case of formaldehyde 
photodissociation, for example, the roaming effect manifests itself 
by a hydrogen atom nearly dissociating and starting to 
orbit the HCO fragment at long distances  
and later returning to abstract the other hydrogen and form the 
products H$_2$ and CO. 
Long-range interactions between 
dissociating fragments  allow the possibility of reorientational dynamics 
that can result in a different set of products 
and/or energy distributions than the one expected from MEP intuition,
while a dynamical bottleneck prevents 
facile escape of the orbiting H atom.

The roaming effect has now been identified in a variety of different 
types of reactions; for example, those involving excited electronic states
\cite{NorthScience2012} or isomerization \cite{Ulusoy13,Ulusoy13b}. 
These studies have identified  some general characteristics of the 
roaming mechanism and point out the need for extending the theories of chemical reactions.   

TST is a fundamental approach to calculating chemical reaction rate constants, 
and can take  various forms, such as RRKM theory \cite{Forst03} or
variational transition state theory (VTST)\cite{Truhlar1984}. 
The central ingredient of TST is the concept of a {\it dividing surface} (DS), 
which is a surface the system must cross in order 
to pass from reactants to products (or the reverse). By its very definition, the DS belongs neither to reactants nor to products 
but is located  at the interface between these two species; this is the essence of the notion of {\it transition state}. Association of 
transition states with saddle points on the potential energy surface (PES) (and their vicinity) has a long history of successful applications in chemistry,
and has provided great insight into reaction dynamics \cite{Levine09,Carpenter84,Wales03}. 
Accordingly, much effort has been devoted to connecting roaming reaction pathways
with the existence (or not) of particular saddle points on the PES, as is evidenced 
by continued discussion of the role of the so-called ``roaming saddle''  
\cite{shepler2011roaming,Harding_et_al_2012}.

It is in fact reasonable to expect that in cases where reactions proceed 
without a clear correlation to saddles of the PES, they
are mediated by transition states that are dynamical in nature, i.e. 
{\it phase space structures}.  Phase space formulations of TST 
have been known since the beginning of the theory \cite{Wigner38}. 
Only in recent years, however, has the phase space (as 
opposed to a configuration space) formulation of TST 
reached conceptual and computational maturity \cite{Wiggins08} for systems with more that two degrees of freedom.
Fundamental to this development is the recognition of the role of phase space objects, namely {\it normally hyperbolic 
invariant manifolds} (NHIMs) \cite{Wiggins_book1994}, in the construction of relevant DSs for chemical reactions.
While the NHIM approach to TST has enabled a deeper understanding of reaction dynamics for systems with many ($\geq3$) 
degrees of freedom (DoF) \cite{Wiggins08,ezra2009microcanonical}, 
its practical implementation has relied strongly on 
mathematical techniques to compute NHIMs, such as the normal form theory 
\cite{Wiggins_book03}. Normal form theory, as 
applied to reaction rate theory, requires the existence of a saddle of index 
$\geq 1$ \cite{Wiggins08} on the PES to construct 
NHIMs and their attached DSs. For dynamical systems with 
two DoF the NHIMs are just unstable periodic orbits (PO), which have long been
known in this context as Periodic Orbit Dividing Surfaces (PODS).
(We recall that a PO is an invariant manifold.  In phase space, an unstable PO forms 
the boundary of the dividing surface for 2 DoF.  For natural Hamiltonian systems, kinetic plus potential energy,
with 2 DoF, the PODS defines a dividing line in configuration space between
reactants and products \cite{Pechukas81}.)
As we shall see, 
these particular hyperbolic invariant phase space structures (unstable POs) 
are appropriate for describing reaction dynamics in 
situations where there is no critical point of the 
potential energy surface in the relevant region of configuration space.

A common characteristic of systems exhibiting 
roaming reactions studied so far is the presence of long range interactions 
between the fragments of the dissociating molecule. 
This characteristic is typical of ion-molecule reactions and 
roaming is clearly expected to be at play in these reactions.
The theory of ion-molecule reactions has a long history going 
back to Langevin \cite{Langevin1905}, who investigated the interaction 
between an ion and a neutral molecule in the gas phase and derived an 
expression for ion-molecule collisional capture rates. 
As researchers have sought to develop  models to account for 
data on ion-molecule reactions, 
there has been much debate in 
the literature concerning  the interpretation of experimental results. 
Some results support a model for reactions taking place via the so-called 
loose  or {\it orbiting transition states} (OTS), while others rather suggest that the reaction operates through a {\it tight transition state} (TTS)
(for a review, see  Ref.~\onlinecite{Chesnavich82}). 
In order to explain this puzzling situation the concept of {\it transition state switching} 
was developed \cite{Chesnavich82}, where both kinds of TS (TTS and OTS) 
are present and determine the reaction rate. 
(See also the unified statistical theory of Miller \cite{Miller76a}.) 
Chesnavich presented a simple model to illustrate these ideas  
\cite{Chesnavich1986}. This relatively simple model 
has all the ingredients required to manifest the roaming effect \cite{MauguiereCPL2014}, and the present
work extends our investigations of the dynamics of Chesnavich's model.

In a recent study \cite{MauguiereCPL2014} we have revisited the Chesnavich model \cite{Chesnavich1986} 
in light of recent developments in TST. We have shown that, for barrierless systems such as ion-molecule reactions, 
the concepts of OTS and TTS can be clearly formulated in terms of well defined phase space geometrical objects 
(for recent work on the phase space description of OTS, see  Ref.~\onlinecite{Wiesenfeld05}). We demonstrated
how OTS and TTS can be identified with well defined {\it phase space dividing surfaces} attached to NHIMs. 
Moreover, this study showed that new reaction pathways, sharing all the characteristics of roaming reactions,  
may emerge and that 
they are associated with free rotor periodic orbits, 
which emanate from {\it center-saddle bifurcations} (CS). 

The Chesnavich model for ion-molecule reactions, parametrised in such a way to represent different classes of molecules with
early and late transition states, offers a useful theoretical
laboratory for investigation of  the evolution of phase space structures relevant to
roaming  dynamics, both in energy and as a function of additional potential function parameters;
such a study is the aim of the present article. 
The birth of new reaction pathways
in phase space associated with non-linear mechanical resonances 
raises questions concerning the applicability of statistical models.
In the present work 
we investigate these questions with a detailed {\it gap time analysis} \cite{Thiele62,Thiele62a,ezra2009microcanonical} 
of direct and roaming dynamics. 

The paper is organised as follows. In subsection \ref{subsec:syst_Hamiltonian} 
we introduce the Hamiltonian of the system to be studied. 
We then proceed to summarise work on the utility of NHIMs 
in the context of TST and discuss the DS associated with them in the subsection 
\ref{subsec:TST}. Subsection \ref{subsec:NHIMs} concludes section \ref{sec:Hamiltonian} 
with a discussion of the application of the NHIM 
approach to the definition of the TTS and OTS in the Chesnavich model. 
Section \ref{sec:roamdyn} presents a dynamical study of the roaming mechanism. 
In subsection \ref{subsec:roaminginterp}, we provide a discussion of the   
roaming phenomenon 
based on an analysis of the dynamics of the Chesnavich model 
for two different values of the parameter that controls the transition state switching in this model.
The role of the so-called ``roaming saddle'' 
in the dynamical interpretation of the roaming phenomenon is also discussed. 
Section \ref{gaptime} provides a  gap time analysis for the Chesnavich model, while 
section \ref{sec:conclusion} concludes.

\newpage

\section{Hamiltonian, NHIMs and transition states}
\label{sec:Hamiltonian}

\subsection{System Hamiltonian}
\label{subsec:syst_Hamiltonian}
More than thirty years ago,  the transition state switching model was proposed 
to account for the competition between multiple transition 
states in ion-molecule reactions
(for a review, see Ref.~\onlinecite{Chesnavich82}). 
Multiple transition states were studied by Chesnavich in the reaction 
CH$_4^+$ $\rightarrow$ CH$_3^+$ + H using a simple model Hamiltonian \cite{Chesnavich1986}. The model system consists of two parts: a rigid, 
symmetric top representing the CH$_3^+$ cation, and a mobile H atom. 
In the following, we employ a simplified version of 
Chesnavich's model restricted to two DoF to study roaming.

The Hamiltonian for planar motion with zero overall angular momentum is:
\begin{equation}
\label{eq:ham}
H = \frac{p_r^2}{2 \mu} + 
\frac{p_{\theta}^2}{2} \left (\frac{1}{I_{CH_3}} + \frac{1}{ \mu r^2} \right ) + V(r,\theta),
\end{equation}
where $r$ is the distance between the centre of mass 
of the CH$_3^+$ fragment and the hydrogen atom. The coordinate 
$\theta$ describes the relative orientation 
of the two fragments, CH$_3^+$ and H, in a  plane. 
The momenta conjugate to these coordinates are $p_r$ 
and $p_{\theta}$, respectively, while $\mu$ is 
the reduced mass of the system and $I_{CH_3}$ is 
the moment of inertia of the CH$_3^+$ fragment. The 
potential $V(r, \theta)$ describes the so-called transitional mode. 
It is generally assumed that in ion-molecule reactions the different modes of the system 
separate into intramolecular (or conserved) and intermolecular (or transitional) 
modes \cite{VandeLinde90a,Peslherbe95,Klippenstein2011}. The potential 
$V(r,\theta)$ is made up of two terms:
\begin{equation}
\label{eq:pot}
V(r,\theta) = V_{CH}(r)+V_{coup}(r,\theta),
\end{equation}
with:
\begin{subequations}
\label{eq:subpot}
\begin{align}
V_{CH}(r) &= \frac{D_e}{c_1-6} \left\{ 2(3-c_2)\exp 
\left[c_1(1-x)\right] 
- (4c_2-c_1c_2+c_1) x^{-6} - (c_1-6)c_2 x^{-4} \right\}, \\
V_{coup} (r,\theta) & = \frac{V_0(r)}{2} \left [ 1-\cos(2\theta) \right ], \\
V_0(r) & = V_e \exp \left [-\alpha(r-r_e)^2 \right ].
\end{align}
\end{subequations}
Here $x=r/r_e$, and parameters fitted to reproduce data from CH$_4^+$ species are: 
dissociation energy $D_e = 47$ kcal/mol and equilibrium distance 
$r_e = 1.1$ \AA.  Parameters $c_1 = 7.37$,  $c_2 = 1.61$, fit the 
polarizability of the H atom and yield a stretch harmonic frequency of 3000 cm$^{-1}$. 
Finally, $V_e = 55$ kcal/mol is the equilibrium barrier height for internal rotation, chosen so that at $r = r_e$ the hindered rotor has, in the low energy 
harmonic oscillator limit, a bending frequency of 1300 cm$^{-1}$. 
The masses are taken to be $m_H = 1.007825$ u, $m_C = 12.0$ u,
and the moment of inertia $I_{CH_3} = 2.373409 $ u\AA$^2$.
The parameter $\alpha$ controls the rate of conversion of the transitional mode from the
angular to the radial mode. By adjusting this parameter one can control whether the conversion occurs `early' or `late' along the reaction coordinate $r$. 
For our study we will study the two cases  $\alpha = 1$ \AA$^{-2}$, which corresponds to a late conversion, and 
$\alpha = 4$ \AA$^{-2}$, which corresponds to an early conversion.

Figs~\ref{fig1}a and \ref{fig2}a show contour plots of the potential function 
as well as representative periodic orbits (see section \ref{sec:roamdyn}) for $\alpha=1$
and  $\alpha=4$, respectively.
In Table~\ref{table:equil}, the stationary points of the potential function for 
the two values of the parameter $\alpha=1$ and $\alpha = 4$ are tabulated and are labelled according 
to their stability properties. The minimum for CH$_4^+$ (EP1) is of center-center (CC) stability type, 
which means that it is stable in both coordinates, $r$ and 
$\theta$. The saddle, which separates two symmetric minima 
at $\theta= 0$ and $\pi$ (EP2), is of center-saddle (CS) type, i.e. stable in $r$ 
coordinate and unstable 
in $\theta$. The maximum in the PES (EP4) is a 
saddle-saddle (SS) equilibrium point. The outer saddle (EP3) is a CS equilibrium point.

\begin{table}[H]
\begin{center}
\begin{tabular}{|c|c|c|c|c||c|c|c|c|c|}
\hline
\multicolumn{5}{|c||}{$\alpha=1$} & \multicolumn{5}{c|}{$\alpha=4$} \\
\hline
E (kcal mol$^{-1}$) &  $r$ (\AA)   &  $\theta$ (rad) & Stability & Label & E (kcal mol$^{-1}$) &  $r$ (\AA)   &  $\theta$ (rad) & Stability & Label \\
\hline
-47.0       & 1.1      &  0            & CC     & EP1   & -47.0        & 1.1      &  0           & CC     & EP1  \\
    8.0       & 1.1      & $\pi/2$   & CS     & EP2   &    8.0        & 1.1      & $\pi/2$  & CS      & EP2   \\
 -0.63   & 3.45   & $\pi/2$   & CS      & EP3   &  -6.44   & 1.96   & $\pi/2$  & CS      & EP3   \\
  22.27 & 1.63   & $\pi/2$   & SS      & EP4   &   8.82 & 1.25   & $\pi/2$  & SS      & EP4   \\
\hline
\end{tabular}
\end{center}
\caption{\label{table:equil} Equilibrium points for the potential $V(r, \theta)$ ($\alpha = 1$ and $4$). 
(CC) means a center-center equilibrium point (EP), (CS) a center-saddle EP and (SS) a saddle-saddle EP.}
\end{table}

The MEP connecting the minimum EP1 with the saddle EP2 at $r=1.1$ \AA  \; (see Figs~\ref{fig1}a and \ref{fig2}a) 
describes a reaction involving `isomerisation' between two symmetric minima. These two isomers cannot of course  
be distinguished physically for a symmetric Hamiltonian. 
The MEP for dissociation to radical 
products (CH$_{3}^{+}$ cation and H atom) follows the line $\theta=0$ with $r \rightarrow \infty$ 
and has no potential barrier (or, one might locate the barrier at infinity). 
Broad similarities between the Chesnavich model and the photodissociation of formaldehyde and other molecules for 
which the roaming reaction has been observed can readily be identified. 
In the Chesnavich model we recognize two reaction 
`channels', one leading to a molecular product 
by passage through an inner TS, and one to radical 
products via dissociation. Moreover, 
a saddle (EP3) exists just below the dissociation threshold, just as has been found in molecules 
showing the roaming effect.

In the remainder of this article we show that,
by adopting a phase space perspective and employing the appropriate transition states 
defined in phase space, not only is the dynamical meaning of the
roaming mechanism revealed but, most importantly, 
this dynamics is shown to be intimately associated with
the generic behavior of non-linear dynamical systems in parameter space,
where bifurcations and resonances may occur and 
qualitatively different dynamics (reaction pathways) are born.

\subsection{Transition states, dividing surfaces and statistical assumptions}
\label{subsec:TST}

TST is based on certain fundamental assumptions \cite{Wigner38,Baer96,Levine09}.  
Once these are  accepted (or tested for the problem considered), TST provides a powerful and very simple
tool for computing the rate constant of a given reaction. 
One of the assumptions is the existence of a DS 
having the property that classical trajectories 
originating in reactants (resp. products) cross this surface 
only once in proceeding to products (resp. reactants). 
Such a DS therefore
separates the phase space into two distinct regions, reactants and products, 
and therefore constitutes the boundary between them.
The definition of the DS given above is fundamentally
dynamical in nature (the local non-recrossing condition).
The DS is in general a surface in the phase space of the system  
under consideration.
Computing, or locating, a phase space surface (hypersurface/volume) 
that realises the first assumption of TST is in general  not an easy task 
as one has to find a codimension one hypersurface in a $2n$-dimensional space for 
an $n$ DoF system. 
As discussed below, the NHIM approach to TST provides a solution to this problem.

A comment on nomenclature:
The term `transition state' is sometimes used to designate 
a saddle point of the system potential energy surface. 
Identification of the \textit{transition state}  
with a point in configuration space is of course 
misleading; a transition state is 
more precisely defined as the manifold  of phase space points 
where the transition between reactants and products occurs. 
The phase space DS defined above is 
just such a collection of phase space transition points.
Confusion between saddle points and TS (DS) 
arises in situations where the system has to overcome
a barrier in the PES in order to react. 
In such a case, there is a saddle point (index one) at the top of the barrier 
and the DS (TS) originates (in phase space) in the vicinity of this saddle point. 
However, there are situations for which the reaction does not proceed via a potential
barrier and in these cases one has to find other phase space structures 
that define DS (TS). 
We have seen that the Chesnavich model provides an example where the outer TS is not associated
with any potential saddle.

In searching for the appropriate DS for which the (local) 
non-recrossing property applies, and thus the minimal reactive flux criterion,  
it is reasonable to start with 
stationary points on the PES.
However, minimization of the flux by varying the DS in configuration space 
as in variational transition state theory (VTST)\cite{Truhlar1984}
may give a better DS. In this approach, the DS is still 
defined in configuration space but its location along some 
reaction path is determined by a variational principle. 
One can also investigate the flux through surfaces 
of specified geometry to determine optimal dividing surfaces in a given family of such surfaces
(for an example of such a surface applied to the roaming phenomenon see Ref.~\onlinecite{Klippenstein2011}). 

The minimal flux through the DS requirement can be cast into a minimum 
of the sum of states at the DS. 
As we move along some reaction coordinate from reactants to products, 
there is two competing effects which affect the sum of states 
in the DS \cite{Miller76a}. First, as we move to the dissociation products the 
potential energy is constantly rising and the available kinetic energy is 
decreasing which has the effect of
lowering the sum of states. The second effect is a 
lowering of the vibrational frequencies at the DS that tends to increase the sum of states. 
These two competing effects result in a minimum in the sum of states 
located at some value of the reaction coordinate. This minimum has been
called an ``entropic barrier'' for the reaction or a tight transition state. 
On the other hand, in the orbiting model of a complex forming the DS 
is located at the centrifugal barrier induced by the effective potential (the 
orbiting TS) \cite{Miller76a,Chesnavich82}. 
In general the TTS and OTS are not located at the same position along the 
reaction coordinate and so one can ask which of these
two DS should be used to compute the rate of the reaction. 
This problem gives rise to the theory of multiple transition states where
one has to decide which DS (TTS or OTS) to use in the computation of the rate of the reaction. 
The Chesnavich model provides 
an excellent example \cite{Chesnavich1981,Chesnavich82,Chesnavich1986}. 
In this model both TS (DS) exist simultaneously and the actual TS (DS) for the
computation of the reaction rate in a naive TST calculation is the one 
giving the minimal flux or, equivalently, the minimal sum of states. 
Millers's approach provides a unified theory approriate when the fluxes
associated with each DS are of comparable magnitude.
We will see in the next paragraph how the Chesnavich model is treated 
within the NHIM approach to TST and in the next section how this approach 
relates to the roaming phenomenon.

The other fundamental assumption of TST is that of statistical dynamics.
If one considers the reaction at a specific energy, 
the statistical assumption requires that throughout the dissociation of the molecule all phase 
space points are equally probable on the timescale of reaction \cite{Baer96}. 
This assumption is equivalent to saying that the redistribution of the energy 
amongst the different DoF of the system on the reactant side of the DS is fast compared to the rate of the reaction,
and guarantees a single exponential decay for the reaction 
(random lifetime assumption for the reactant part of the phase space \cite{Thiele62}). 
In section~\ref{gaptime} we will investigate this statistical assumption 
for the roaming phenomenon by studying the gap time distributions in the ``roaming region'' 
defined in section~\ref{sec:roamdyn}.

\subsection{Normally hyperbolic invariant manifolds and their related dividing surfaces}
\label{subsec:NHIMs}

In this section we summarise recent work which has 
shown the usefulness of NHIMs in the context of TST \cite{Wiggins08}.
In the previous subsection we recalled that TST is
build on the assumption of the existence of a DS separating the phase space
into two parts, reactant and products. The construction of this surface has been the subject of many studies. 
As we emphasized, the DS is in general  a surface in phase space, 
and the construction of such surfaces for systems with three 
and higher DoF has until recently been a major obstacle in the development of the theory.

For systems with two DoF described by a natural Hamiltonian, kinetic plus potential energy, 
the construction of the DS is relatively straightforward. 
This problem was solved during the 1970s by McLafferty, Pechukas and Pollak 
\cite{Pechukas73,Pechukas77,Pollak78,Pechukas79}. 
They showed that the DS at a specific energy is intimately related to an invariant phase 
space object, an unstable PO.
The PO defines the bottleneck in 
phase space through which the reaction occurs 
and the DS which intersects trajectories evolving from reactants to products can be shown
to have the topology of a hemisphere whose boundary is the PO 
\cite{waalkens2004direct,wiggins2001impenetrable}. The same 
construction can be carried out for a DS intersecting trajectories travelling 
from products to reactants and these two
hemispheres form a sphere for which the PO is the equator.

Generalisation of the above construction to higher dimensional systems has been a major 
question in TST and has only received a 
satisfactory answer relatively recently
\cite{waalkens2004direct,wiggins2001impenetrable}. 
The key difficulty 
concerns the higher dimensional analogue of the unstable PO used 
in the two DoF problem for the construction of the DS.
Results from dynamical systems theory show 
that 
transport in phase space is controlled by various high dimensional manifolds,
Normally Hyperbolic Invariant Manifolds (NHIMs), which are the natural generalisation of the unstable
PO of the two DoF case. Normal hyperbolicity of
these invariant manifolds means that they are, in a precise sense, structurally stable,
and possess stable and unstable invariant manifolds that govern the transport in phase
space \cite{Fenichel1971,Fenichel1977,Fenichel1974,Wiggins_book1994}.

Existence theorems for  NHIMs are well established \cite{Fenichel1971,Fenichel1977,Fenichel1974,Wiggins_book1994},
but for concrete examples one needs methods
to compute them. One approach 
involves a procedure based on Poincar\'e-Birkhoff normalisation: 
the idea is to find a set of canonical coordinates 
by means of canonical transformations that put the 
Hamiltonian of the system in a ``simple'' form in a 
neighbourhood of an equilibrium point of saddle-centre$\cdots$-centre
type (an equilibrium point at which the linearized vector field has one pair of real eigenvalues and $n-1$ imaginary eigenvalues 
for a system of $n$ DoF). The ``simplicity'' comes from the fact that, 
under non-resonance conditions among the imaginary frequencies 
at the saddle point, one can construct an integrable
system valid in the neighbourhood of the equilibrium point and 
thereby describe the dynamics in this neighbourhood very simply. 
With this new Hamiltonian, the geometrical structures that govern reaction dynamics are revealed. 
For two DoF systems, the NHIM is simply a PO.
For an $n > 2$ DoF system at a fixed energy, the NHIM has the topology of a $(2n-3)$ sphere. 
This $(2n-3)$-dimensional sphere is the equator of a $(2n-2)$-dimensional sphere which constitutes the DS. 
The DS divides the $(2n-1)$-dimensional energy surface into 
two parts, reactants and products,
and one can show that it is a surface of {\it minimal flux} \cite{waalkens2004direct}.

The NHIM approach to TST consists of constructing DSs for the reaction studied built from NHIMs,
and constitutes a rigorous 
realisation of the local non-recrossing property. 
Once these geometrical objects (NHIM and DS) are computed  the reactive flux from reactant to products 
through the DS can easily be expressed as the integral of a flux form over the DS. 
Furthermore, it is possible to sample the DS and use this knowledge to propagate 
classical trajectories initiated at the TS (DS).
As noted above,
for the two DoF case, the unstable PO used in the construction of the DS is 
just an example of such a NHIM.
In this paper,
we are concerned with an $n=2$ DoF problem and therefore the NHIMs 
we will be interested in are POs.
Extension to $n >2$ DoF systems is in principle conceptually straightforward.

\newpage

\section{Roaming dynamics}
\label{sec:roamdyn}

\subsection{Dynamical interpretation of the roaming mechanism}
\label{subsec:roaminginterp}

\begin{figure}[H]
\centering
\includegraphics[scale=0.7]{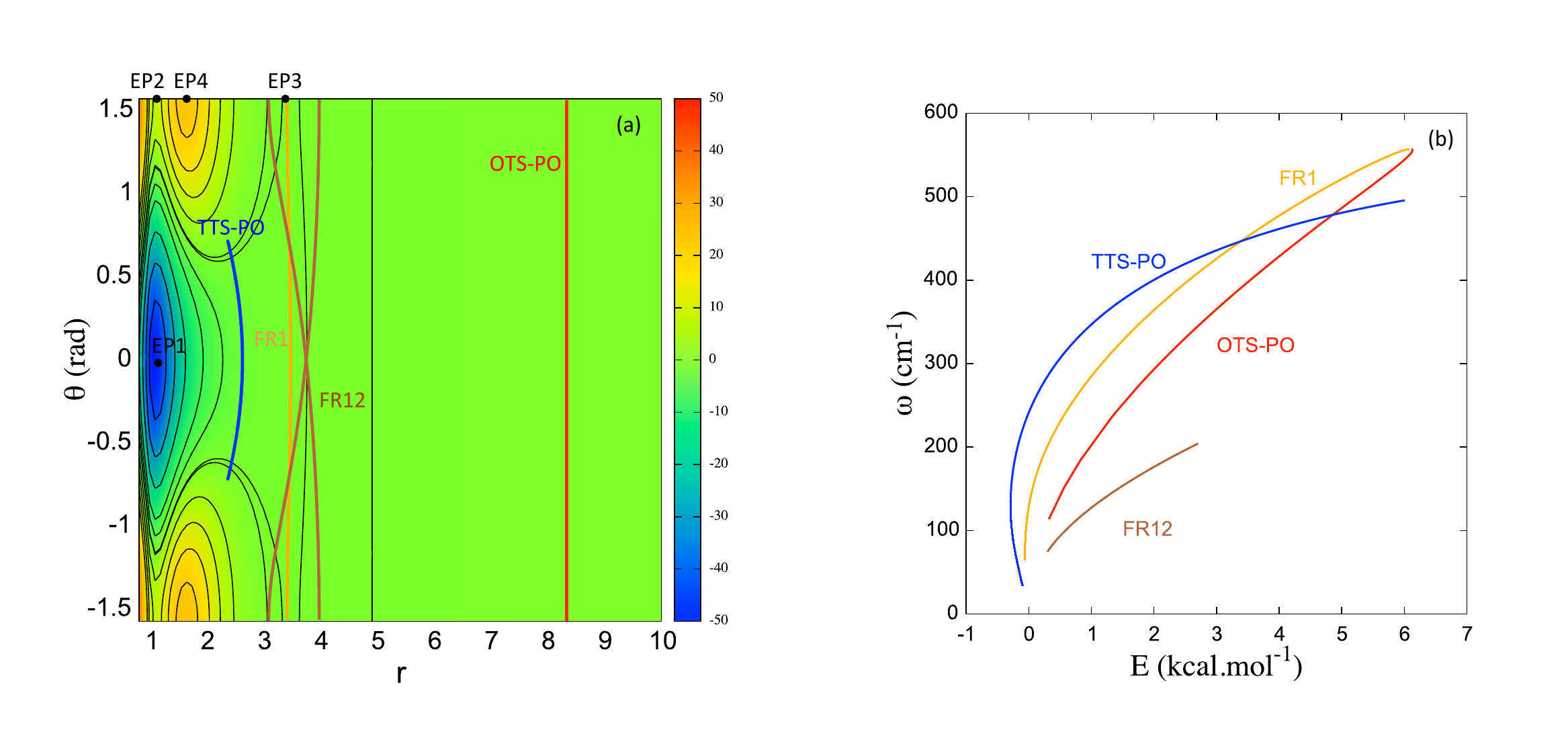}
\caption{(a) Contour plot of the PES for $\alpha=1$ with representative POs. 
(b) Continuation/bifurcation diagram of families of periodic orbits for $\alpha=1$.}
\label{fig1} 
\end{figure}

\begin{figure}[H]
\centering
\includegraphics[scale=0.7]{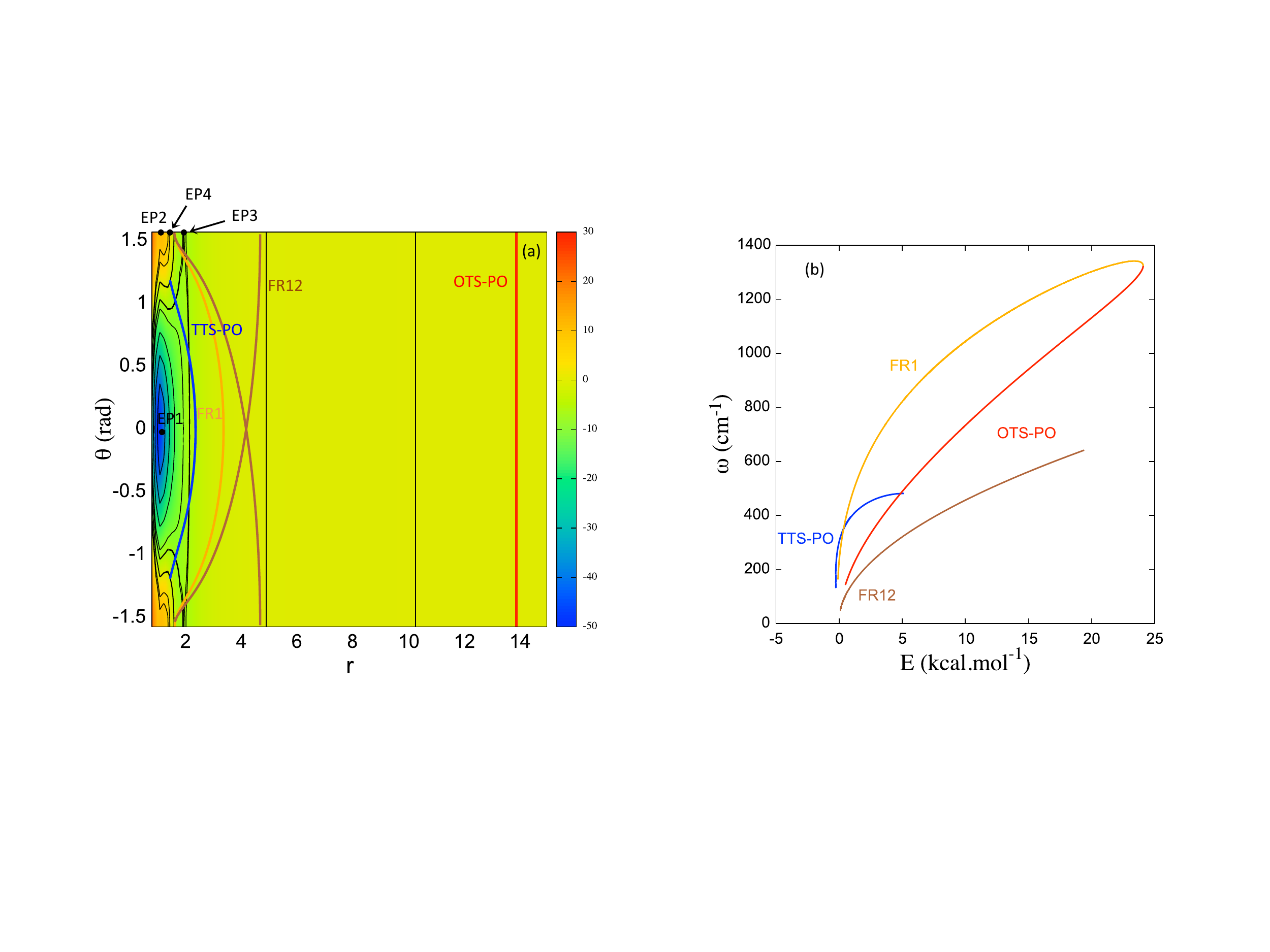}
\caption{(a) Contour plot of the PES for $\alpha=4$ with representative POs. 
(b) Continuation/bifurcation diagram of families of periodic orbits for $\alpha=4$.}
\label{fig2} 
\end{figure}

In the previous section, we discussed the notions of TTS and OTS 
in the context of a reaction occurring without a potential barrier. 
We also discussed
how the NHIM approach to TST provides a rigorous way of constructing a 
DS that satisfies the local no-recrossing requirement of TST. 
To define DSs that are relevant
for the description of reactions in the model Hamiltonian 
defined in Eq.~\eqref{eq:ham}, we need to locate unstable POs.
Periodic orbits for conservative Hamiltonian systems exist in
\emph{families}, where the POs in a family depend on system parameters.
In molecular systems, for example, it 
is very common to consider PO families obtained by variation of the energy of the system 
(see for
example Refs.~\onlinecite{Farantos09} and \onlinecite{Mauguiere10}). 
At critical parameter values (the energy, for example)
bifurcations take place and new families are born. 
Continuation/bifurcation (CB) diagrams are obtained by plotting a PO property as a 
function of the parameter. One important kind of elementary bifurcation is the center-saddle, 
which turns out to be ubiquitous in non-linear dynamical systems \cite{Wiggins_book03}. 
Although periodic orbits, being one dimensional objects, cannot 
reveal the full structure of phase space, they do provide a ``skeleton' 
around which more complex structures such as invariant tori 
develop. Numerous explorations of non-linear dynamical systems by construction of PO 
CB diagrams have been made. 
In particular, for molecules with multidimensional, 
highly anharmonic and coupled potential functions, software has been developed to locate POs
based on multiple shooting algorithms \cite{Farantos98}, 
and has successfully been applied to realistic models of small polyatomic molecules \cite{Farantos09}. 
In Figs.~\ref{fig1}b and \ref{fig2}b such CB diagrams are 
shown for the Chesnavich model for the values of the parameter $\alpha=1$ and
$\alpha=4$ respectively. 
Not all principal families of POs generated from all equilibria are shown, but only those 
which are relevant for our discussion of the roaming phenomenon. 

We identify the DS constructed from the PO denoted TTS-PO in Figs.~\ref{fig1}b and \ref{fig2}b with 
the TTS. These periodic orbits
show hindered rotor behavior. The OTS is related to the centrifugal barrier appearing
due to the presence  of the centrifugal ($\approx r^{-2}$) term 
in the kinetic energy, Eq.~\eqref{eq:ham}. 
There is in fact a PO
associated with the centrifugal barrier, 
referred to as a {\it relative equilibrium}. 
In Figs.~\ref{fig1}b and \ref{fig2}b we refer to  this PO  as OTS-PO 
and we identify the DS constructed from this PO with the OTS. These relative equilibria POs and higher dimensional analogues 
have been studied by Wiesenfeld et \textit{al.} \cite{Wiesenfeld05} 
in the context of capture theories of reaction rates.

We have therefore clearly identified the notions of TTS and OTS found in the literature 
with DSs constructed from NHIMs. 
These TSs (DSs) are surfaces which satisfy rigorously 
the requirement of local no-recrossing TST theory. 
These two TSs (DSs) exist simultaneously for our model Hamiltonian
and in the following we discuss the dynamical consequences 
of this fact and how one can interpret roaming phenomenon in this setting.

\subsubsection{Roaming and non-linear mechanical resonances}
\label{subsec:roaming}

The TTS and OTS are associated with different reactive bottlenecks
in the system, and hence, in a certain sense, with different `reaction pathways'. 
In order to completely dissociate to CH$_3^+$+H, 
the system has to cross the OTS. This surface satisfies a global non-recrossing condition 
(as opposed to a local non-recrossing condition) 
in the sense that once the system crosses this surface in the outward sense
the orbital momentum is an approximate constant of motion (for sufficiently large $r$), 
and the system enters an uncoupled  free rotor regime. 
The TTS, on the other hand, is the DS associated with formation of  
the cation CH$_4^+$. These two DS delimit an intermediate region 
defining the association complex CH$_3^+\cdots$H. 
The picture here is similar to that discussed by Miller 
in his Unified Statistical Theory \cite{Miller76a}, where 
a modified statistical theory is developed to describe
association/dissociation dynamics in the presence of a complex.
In Miller's theory, the complex was associated with a well in the PES, 
whereas in our case, there is actually no potential well 
in the intermediate region between the TTS and OTS 
with which the complex can be unambiguously associated. 
Instead, there are non-linear 
mechanical resonances, which create `sticky' regions in phase space 
(for rigorous results on the notion of stickiness in Hamiltonian systems see \cite{pw94,mg95b}). 
These resonances are marked by the families of periodic orbits 
FR1 and its period doublings (for example FR12), so that 
the phase space region delimited by 
the TTS and the OTS can be thought of as a ``dynamical complex''.

The two TTS and OTS form two phase space bottlenecks between
which trajectories can be trapped for arbitrary long times. 
This trapping is responsible for the existence of trajectories 
for which the hydrogen atom winds around the CH$_3^+$ fragment and ``roams'' 
before exiting the dynamical complex, 
either to reform CH$_4^+$ or to dissociate to CH$_3^+$+H. 
Hence, the dynamical complex defines the \textit{roaming region}.
To study this trapping phenomenon we initiate classical trajectories
on the OTS and follow them either until the 
CH$_4^+$ cation is formed or dissociation back to CH$_3^+$+H occurs. 
Trajectory propagation is then stopped shortly after crossing of the TTS or OTS occurs. 
We make such calculations for two different values of 
the parameter $\alpha$ 
which controls the location of the TS in the Chesnavich transition state switching model. 
In the next paragraph we describe these classical trajectory simulations.

\subsubsection{Classical trajectory simulations}
\label{subsec:cltrajsimul}

To perform our classical trajectory simulation we uniformly sample
trajectory initial conditons on the OTS at constant energy (microcanonical sampling). 
As explained in section \ref{sec:Hamiltonian}, 
the OTS-DS is composed of two parts: one hemisphere for 
which the trajectories cross from reactants to products (forward hemisphere)
and the other for which the trajectories cross from products to reactants 
(backward hemisphere). For the OTS, if we define as reactants the
complex CH$_3^+\cdots$H and as products CH$_3^+$+H, 
in our simulation we are interested only in trajectories lying on the 
backward hemisphere of the DS. We sample  this hemisphere uniformly 
and numerically integrated  the equations of motion 
until the trajectories cross either the OTS (forward hemisphere this time) 
or the TTS (backward hemisphere if CH$_4^+$ is defined as 
reactants and the complex CH$_3^+\cdots$H as the products for the TTS).

We wish to classify trajectories according to qualitatively different types of 
behavior, i.e.,  trajectories associated with different reactive events. 
Two obvious qualitatively different types of trajectories can be identified. 
First, there are trajectories which cross the TTS and form CH$_4^+$. 
These trajectories are `reactive' trajectories. 
Second, there are trajectories which recross 
the OTS to form CH$_3^+$+H. These trajectories are `non reactive'. 

Our classification scheme requires a precise definition of `roaming' trajectories. 
In a previous publication \cite{MauguiereCPL2014} we proposed a classification of 
trajectories according to the number of turning points 
in the $r$ coordinate. 
Here, in light of subsequent investigations involving gap times (see the next section),
we refine this definition of roaming. 
In the present system,
roaming is intuitively associated with motions in which the hydrogen atom orbits the
CH$_3^+$ fragment while undergoing oscillations in the $r$ coordinate. 
For such motions to occur, energy must 
be transferred from the radial to  the angular mode and (see below)
the mechanism for such an energy transfer involves non-linear resonances, which are
manifest by the appearance of the FR1 POs. 
Just as we construct DS associated with the TTS-POs and the OTS-POs, 
it is possible to define a DS associated with the FR1 PO, 
which we denote the FR1-DS. 
To exhibit roaming character according to our revised definition, a trajectory
must cross  the FR1-DS several times. 
Such a trajectory will therefore involve exchange of energy between the radial and angular DoF 
before finding its way to a final state (either CH$_4^+$ or CH$_3^+$+H).

We now define the four categories  of trajectories used in our analysis of the Chesnavich model:
\begin{itemize}

\item Direct reactive trajectories: these trajectories cross  the FR1-DS 
only once before crossing the TTS to form CH$_4^+$.

\item Roaming reactive trajectories: these trajectories cross the FR1-DS 
at least three times  before crossing the TTS to form CH$_4^+$. 
Note that a reactive trajectory has to cross the FR1-DS an odd number of times.

\item Direct non reactive trajectories: these trajectories cross  the FR1-DS 
only twice before crossing the OTS to form CH$_3^+$+H.

\item Roaming non reactive trajectories: these trajectories cross  the FR1-DS 
at least four times before crossing the OTS to form CH$_3^+$+H. 
Note non reactive trajectories have to  cross the FR1-DS an even number of times.

\end{itemize}

Note that, in principle  there may be non reactive trajectories which never cross the FR1-DS 
but which return immediately to recross the OTS. 
The existence of these trajectories is perfectly conceivable as 
the stable and unstable manifolds of
the period doubling bifurcated orbits of the FR1 family, 
could `reflect back' the incoming trajectories. 
However, we find no such trajectories  in our simulations.

Trajectories were propagated and classified into the four different classes 
according to the definitions given above 
for two different values of the parameter $\alpha$ of the Hamiltonian. 
In the Chesnavich model, this parameter controls 
the ``switching'' of the transition state from late to early. 
The switching model was developed in the context of variational TST 
where it is necessary to determine the optimal transition state to 
use in a statistical theory in order to compute the reaction rate. 
A variational criterion is used to select the relevant TS,  tight or loose. 
In the system under study here, 
two phase space DS (TTS and OTS) exist simultaneously, 
and in order to analyze the roaming phenomenon in dynamical terms, both must 
be taken into account. 
In the next section, we investigate the question of the assumption of
statistical dynamics in the roaming region.

\begin{figure}[H]
\centering
\includegraphics[scale=0.3]{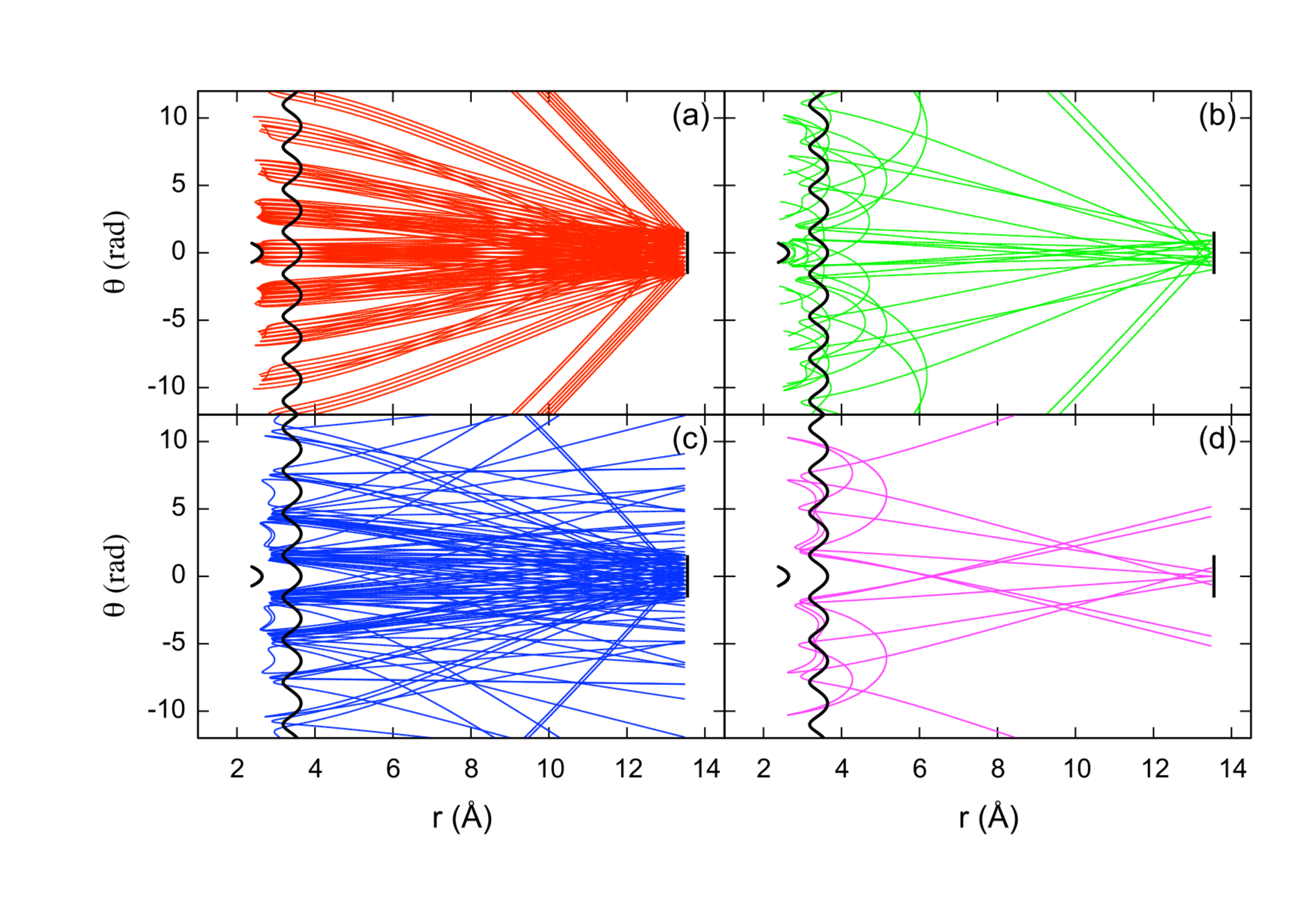}
\caption{The four different types of trajectories for the case $\alpha=1$. 
The thick black curves correspond to the TTS-PO, FR1 PO and
OTS PO, respectively.
(a) direct reactive trajectories (red). (b) Roaming reactive trajectories (green). 
(c) Direct non reactive trajectories (blue). (d) Roaming non
reactive trajectories (magenta).}
\label{fig3} 
\end{figure}

\begin{figure}[H]
\centering
\includegraphics[scale=0.6]{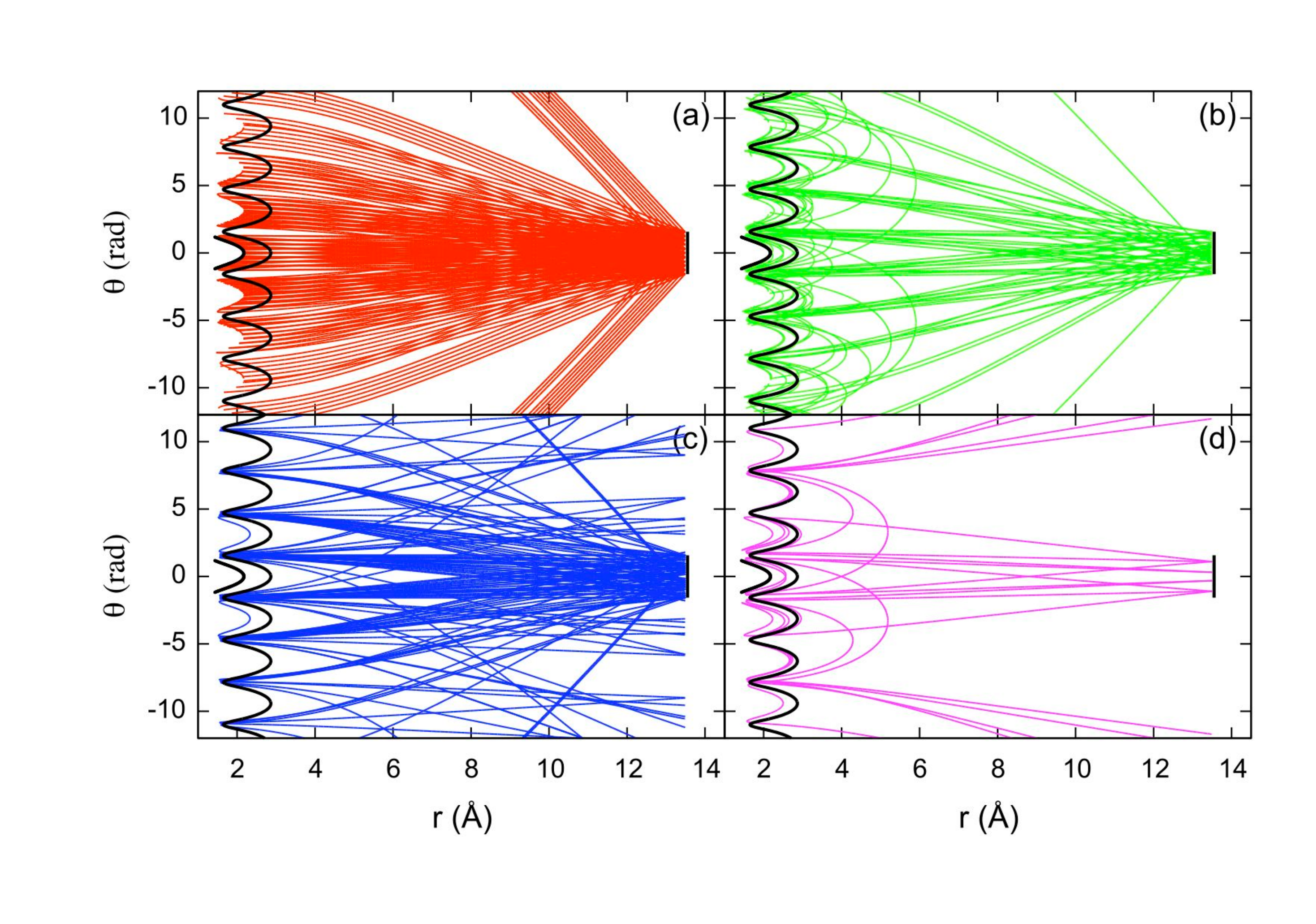}
\caption{The four different types of trajectories for the case $\alpha=4$. 
The thick black curves correspond to the TTS-PO, FR1 PO
and OTS PO, respectively.
(a) Direct reactive trajectories (red). (b) Roaming reactive trajectories (green). 
(c) Direct non reactive trajectories (blue). (d) Roaming non
reactive trajectories (magenta).}
\label{fig4} 
\end{figure}

In Fig.~\ref{fig3} we show the result of our classical trajectory simulations 
at energy $E=0.5$ kcal/mol for the case $\alpha=1$. 
The case $\alpha=1$ corresponds 
to the switching occurring late, which means that if one were to use variational 
TST an OTS would be used to compute the rate. In Fig.~\ref{fig3} the TTS-PO, the FR1 and the 
OTS-PO are represented as thick black curves. 
Each panel of the figure shows trajectories belonging 
to different classes of trajectories that we defined earlier.
Fig.~\ref{fig3}a shows the direct reactive trajectories, Fig.~\ref{fig3}b the roaming reactive trajectories,
Fig.~\ref{fig3}c the direct non reactive trajectories and Fig.~\ref{fig3}d the roaming non reactive trajectories. 
Similarly, Fig.~\ref{fig4} shows the results for the case $\alpha=4$ at the
same energy.

\begin{figure}[H]
\centering
\includegraphics[scale=0.45]{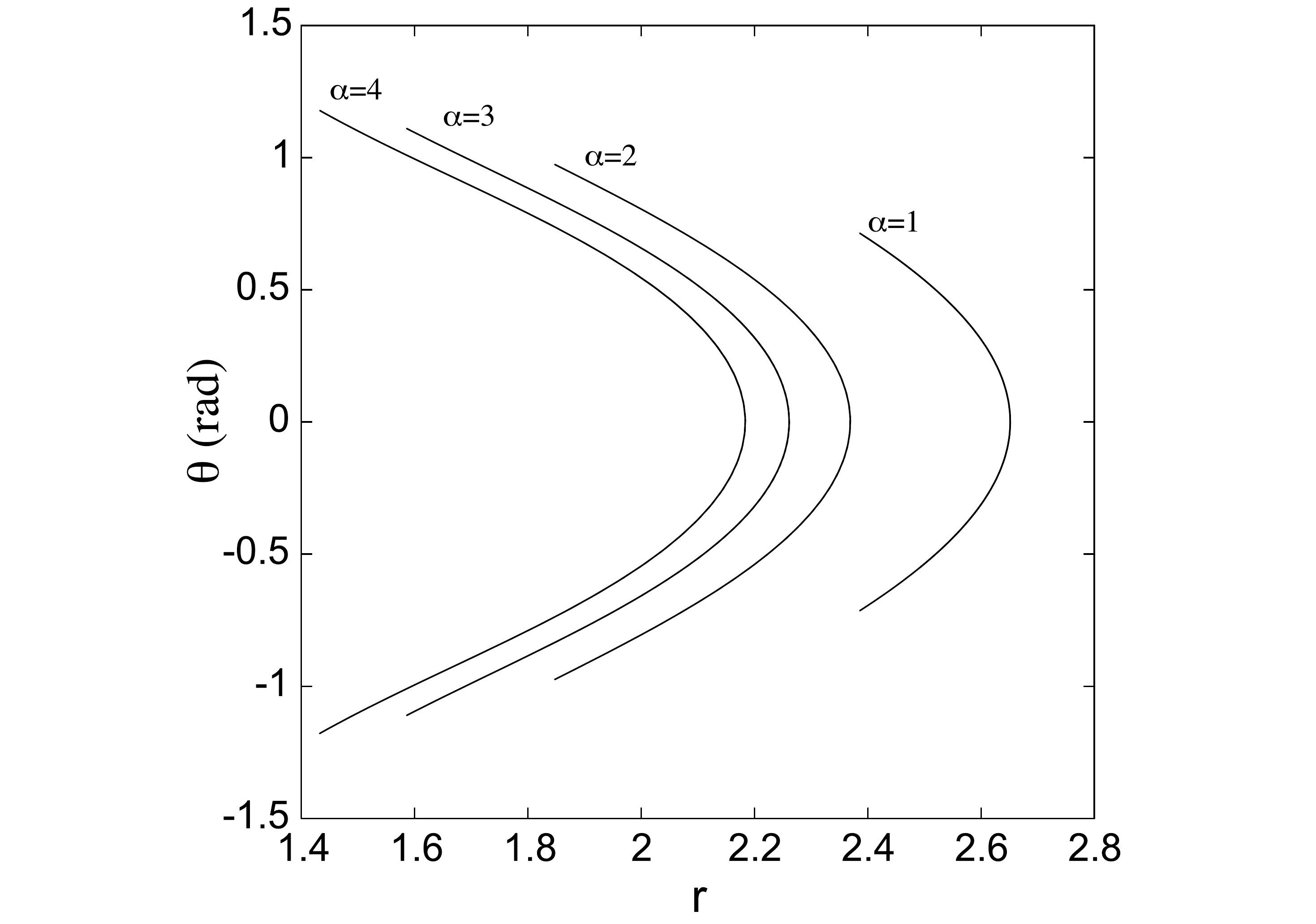}
\caption{Evolution of the TTS-PO with parameter $\alpha$.
Constant energy of 0.5 kcal.mol$^{-1}$}
\label{fig5} 
\end{figure}

\begin{figure}[H]
\centering
\includegraphics[scale=0.5]{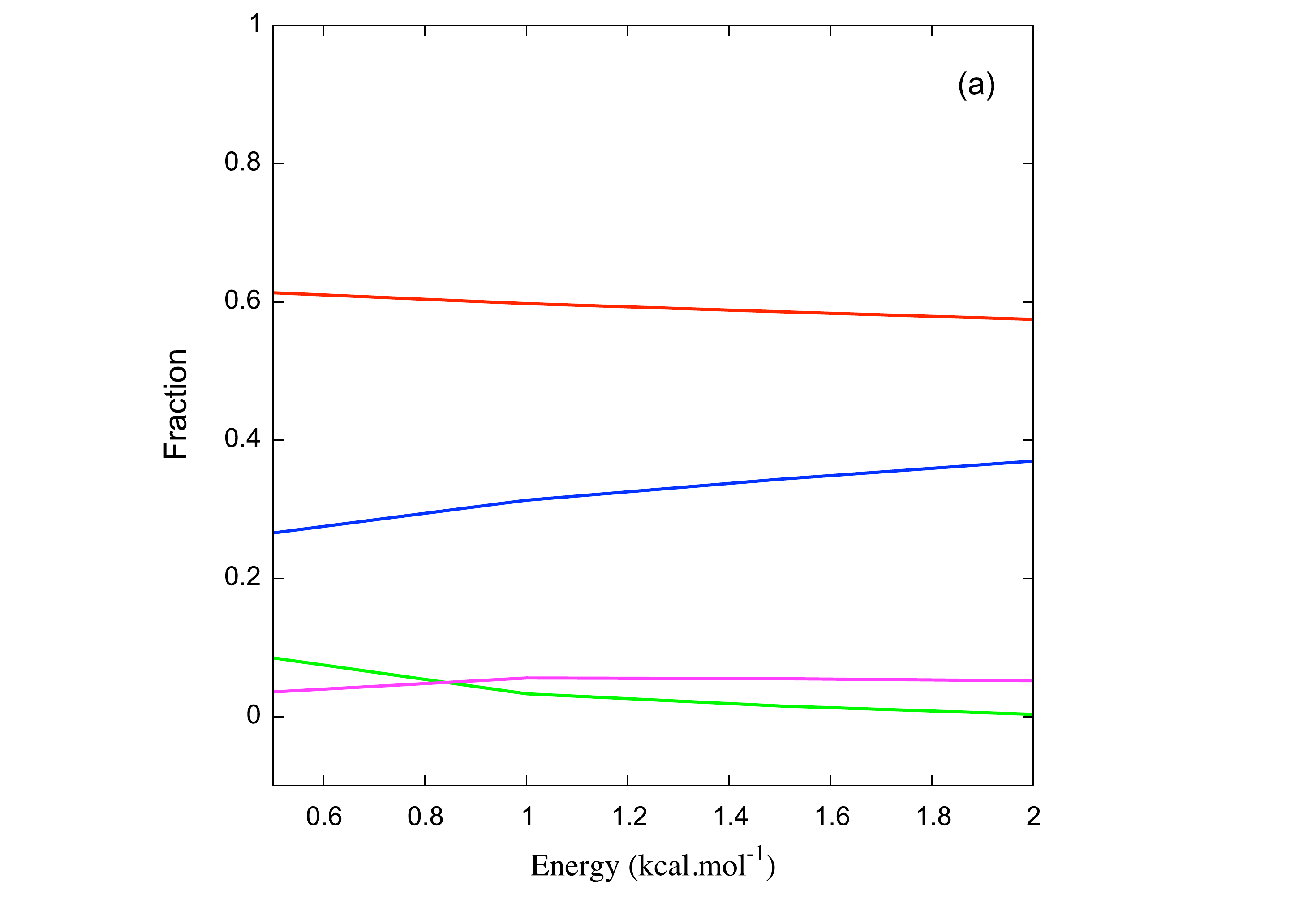}
\includegraphics[scale=0.5]{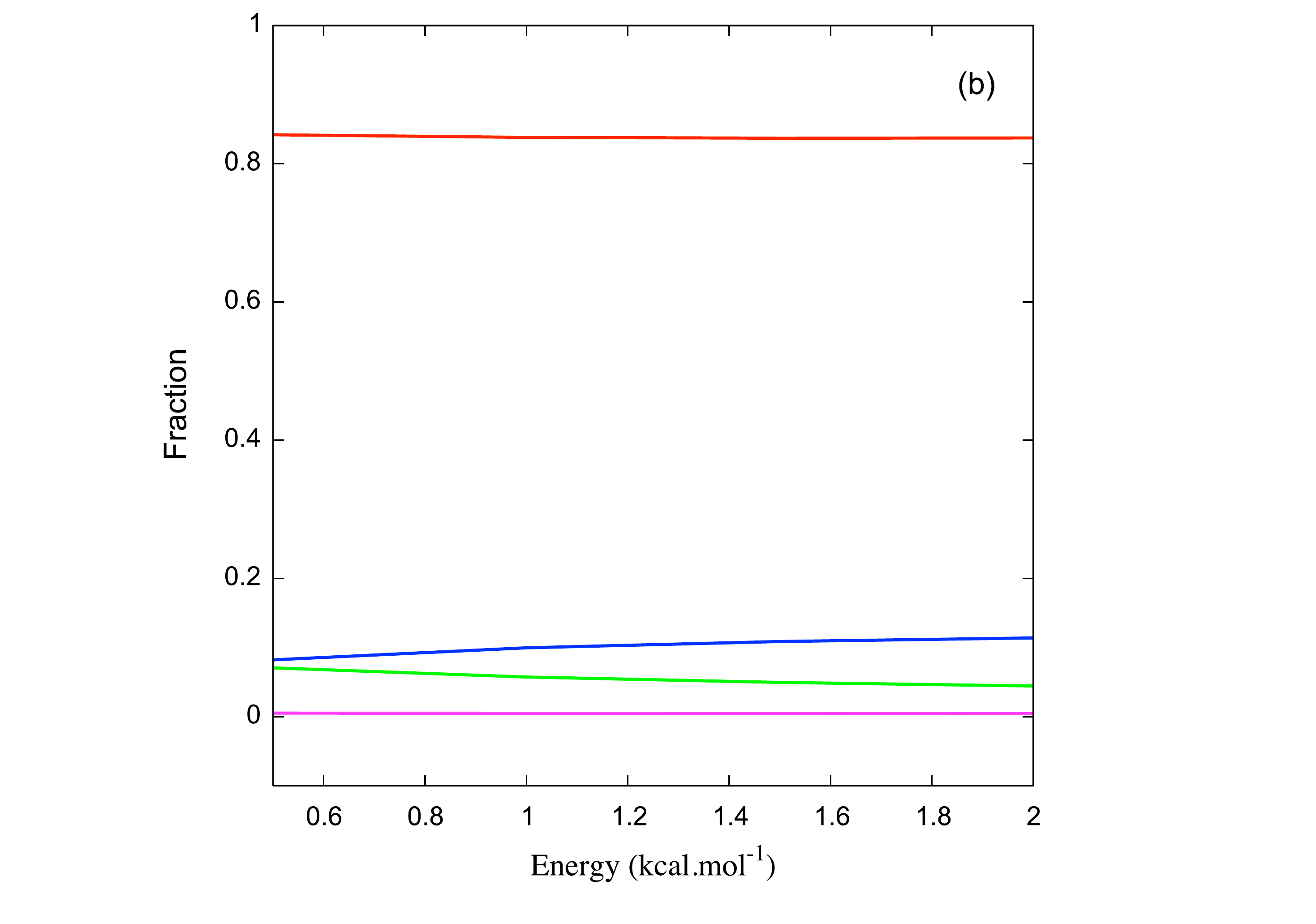}
\caption{(color online) Fractions of different types of trajectories versus energy. Red line is the fraction of direct reactive trajectories, green
for roaming reactive trajectories, blue for direct non reactive and magenta for roaming non reactive trajectories. 
(a) $\alpha=1$, (b) $\alpha=4$.}
\label{fig6} 
\end{figure}

\begin{figure}[H]
\centering
\includegraphics[scale=0.7]{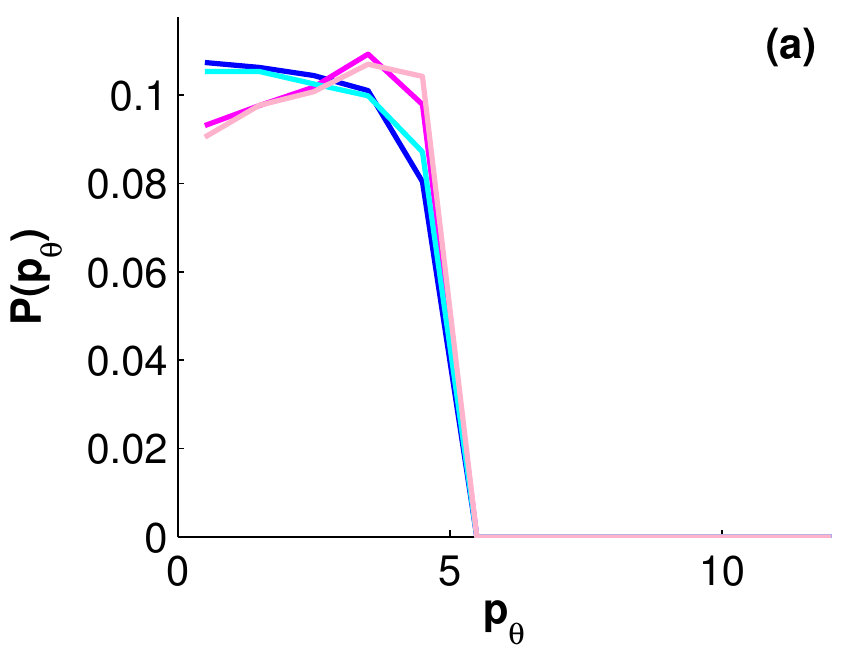}
\includegraphics[scale=0.7]{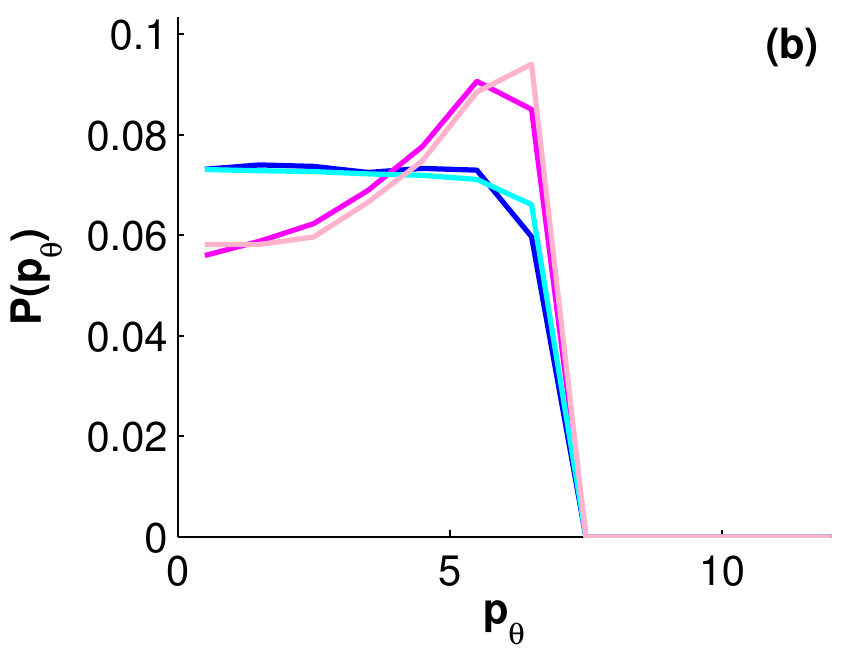}
\includegraphics[scale=0.7]{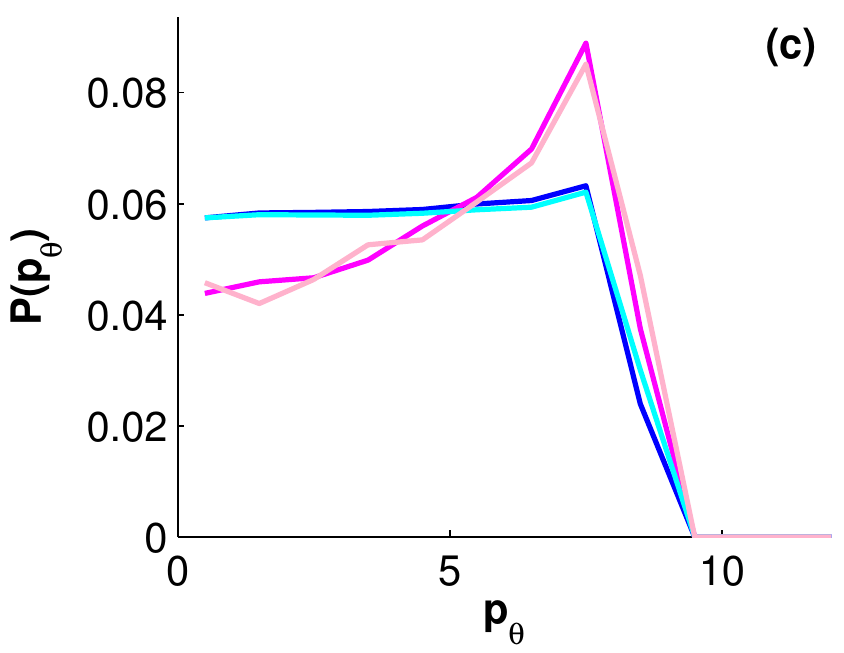}
\includegraphics[scale=0.7]{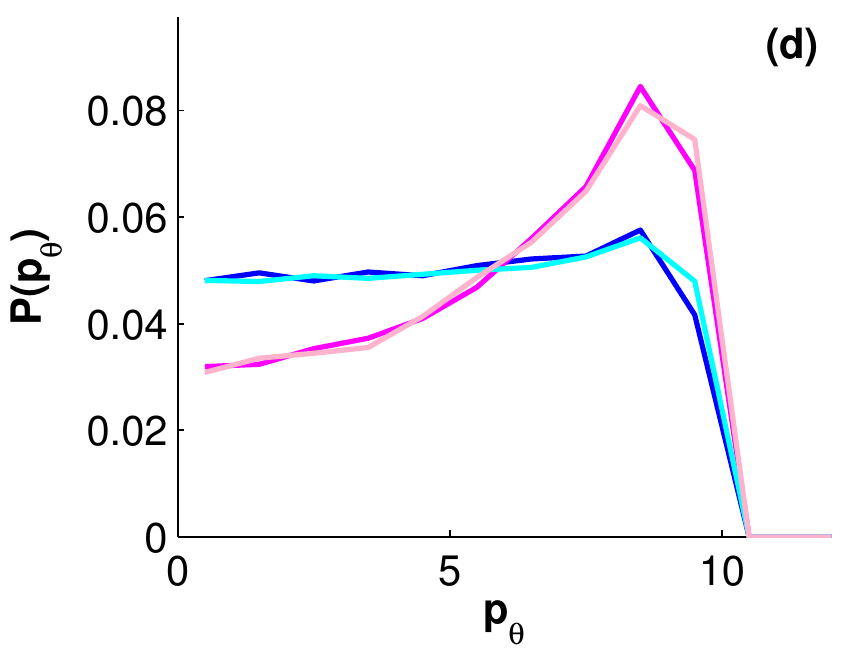}
\caption{Initial and final normalised $p_{\theta}$ distributions for $\alpha=1$. Blue and magenta curves represent
initial distributions for direct non reactive and roaming non reactive trajectories, respectively, and cyan and
pink curves represent the final distributions for direct non reactive and roaming non reactive trajectories, respectively.
(a) Energy E=0.5 kcal.mol$^{-1}$. (b) E=1.0 kcal.mol$^{-1}$. 
(c) E=1.5 kcal.mol$^{-1}$. (d) E=2.0 kcal.mol$^{-1}$.}
\label{fig7} 
\end{figure}

\begin{figure}[H]
\centering
\includegraphics[scale=0.7]{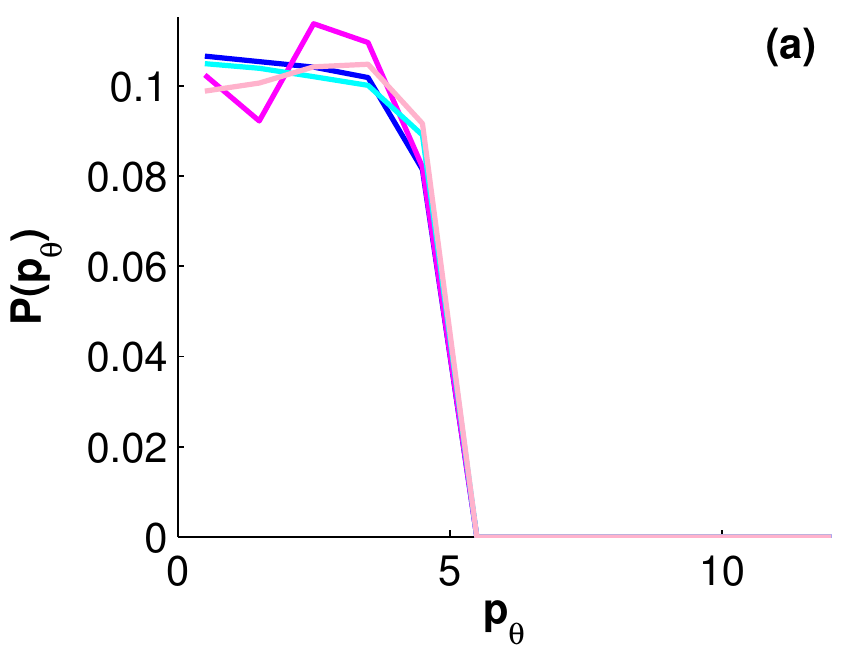}
\includegraphics[scale=0.7]{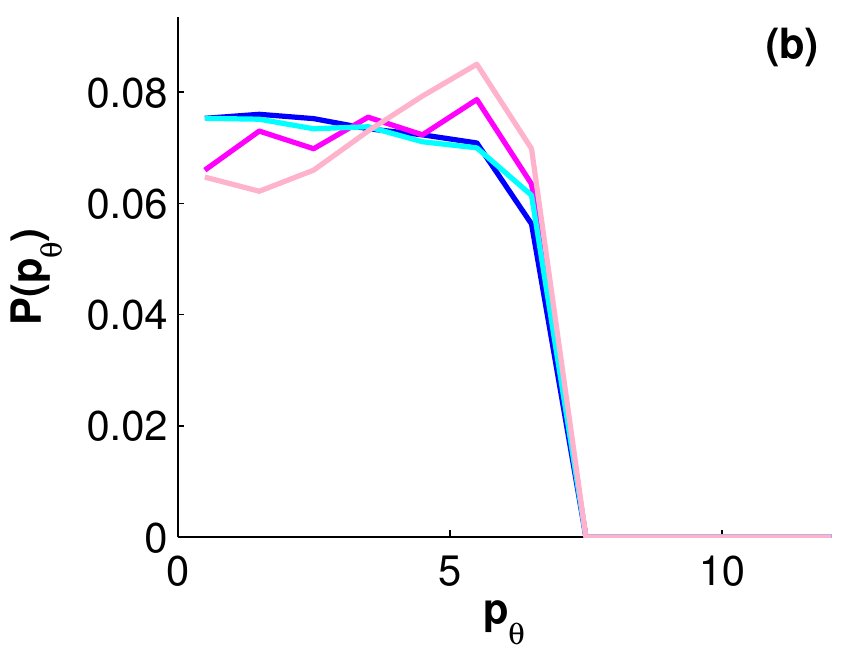}
\includegraphics[scale=0.7]{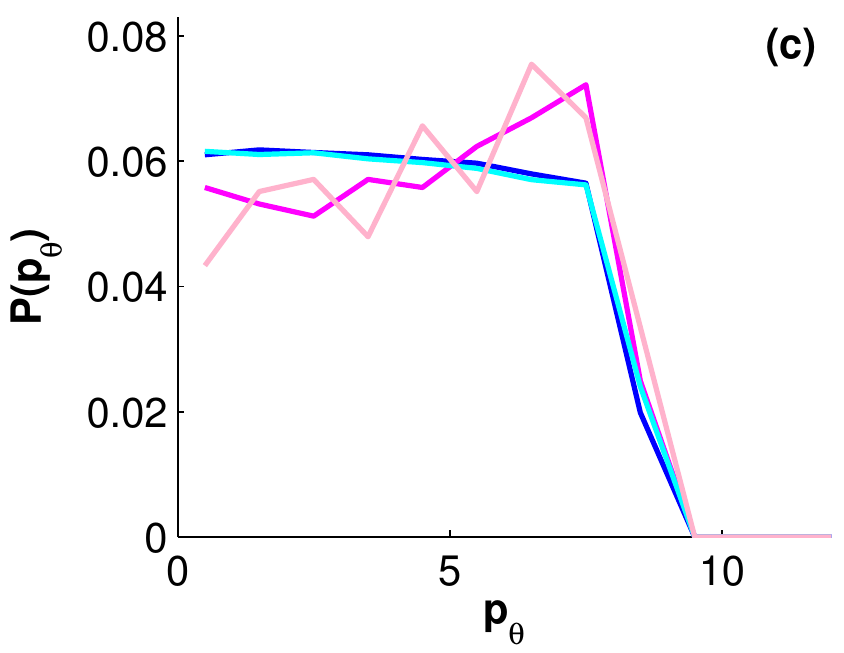}
\includegraphics[scale=0.7]{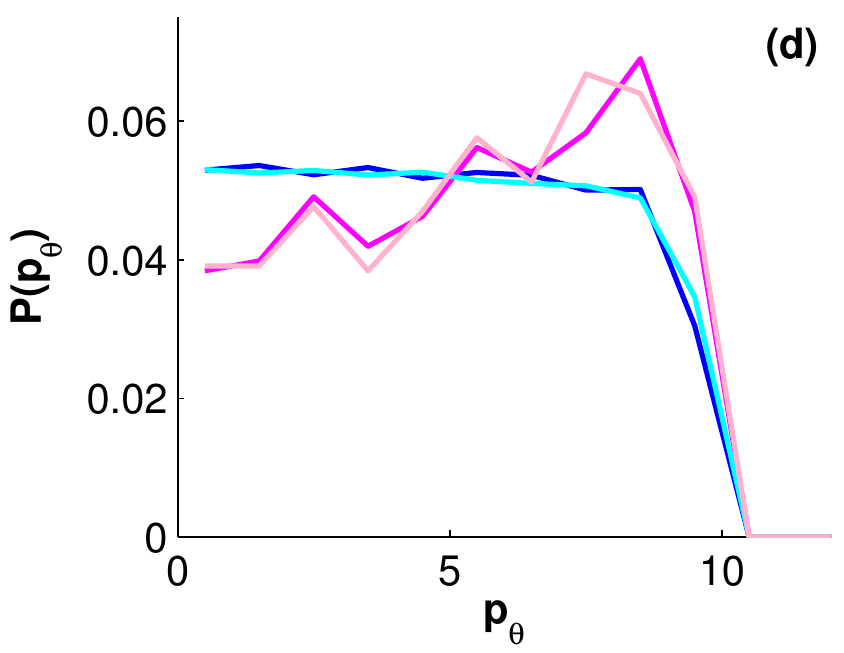}
\caption{Initial and final normalised $p_{\theta}$ distributions for $\alpha=4$. Blue and magenta curves represent
initial distributions for direct non reactive and roaming non reactive trajectories, respectively, and cyan and
pink curves represent the final distributions for direct non reactive and roaming non reactive trajectories, respectively.
(a) Energy E=0.5 kcal.mol$^{-1}$. (b) E=1.0 kcal.mol$^{-1}$. 
(c) E=1.5 kcal.mol$^{-1}$. (d) E=2.0 kcal.mol$^{-1}$.}
\label{fig8} 
\end{figure}

Figure~\ref{fig5} shows the evolution of the TTS-PO with $\alpha$ at constant energy of 0.5 kcal.mol$^{-1}$. 
The location of the OTS remains practically unchanged with $\alpha$. 
In order to quantify the roaming effect we plot in Fig.~\ref{fig6} 
the fractions of the different classes of trajectories versus energy. 
Fig.~\ref{fig6}a is for the case $\alpha=1$ and 
Fig.~\ref{fig6}b for $\alpha=4$. 
It is also instructive to look at the rotor angular momentum ($p_{\theta}$)
distributions for the direct and roaming non reactive
trajectories at the beginning and the end of the trajectory propagation. 
The angular momentum distributions for direct trajectories 
and those exhibiting roaming are found to be qualitatively different in experiments 
\cite{townsend2004roaming} and 
Figs.~\ref{fig7} and \ref{fig8} show that our classification scheme captures
this aspect of the roaming phenomenon for our model system.
Fig.~\ref{fig7} shows initial and final angular momentum distributions for the case $\alpha=1$ at several energies and 
similarly Fig.~\ref{fig8} for the case $\alpha=4$. 
Both initial and final distributions are identical  within the statistical
errors, as is expected since 
the OTS PO defines the only entrance (exit) portal 
for the association (dissociation) of radical reactants (products) to occur.

There has been an interesting discussion in the literature 
concerning the possible existence of a saddle in the PES responsible for the roaming reaction, 
often referred to  as the ``the roaming saddle'' \cite{Harding07,shepler2011roaming,Klippenstein2011,Harding10}. 
Indeed, for the Chesnavich model, there exists such a saddle point on the PES, 
labelled EP3 in Table \ref{table:equil}, 
which could be considered to be a roaming saddle. 
However, as has already been pointed out, transition states 
are in general not associated with particular potential saddles. 
We have shown that the TTS and OTS are the 
dividing surfaces associated with unstable periodic orbits, 
those of TTS-PO and OTS-PO families, 
which originate from 
center-saddle bifurcations (see Figs. \ref{fig1}b and \ref{fig2}b). 
On the other hand, equilibria of the PES, both stable and unstable, 
are essential in tracing the birth of time invariant objects in phase space, 
such as principal families of periodic orbits, tori, NHIMs and 
other invariant manifolds. 

The existence of  ubiquitous center--saddle bifurcations of periodic orbits 
is supported by the Newhouse theorem \cite{Newhouse79,Wiggins_book03},
which was
initially proved for dissipative dynamical systems, and  later extended to 
Hamiltonian systems \cite{Duarte99,Gonchenko00}. 
The theorem states that tangencies of the stable and unstable manifolds 
associated with unstable equilibria and periodic orbits 
generate an infinite number of period doubling and center--saddle bifurcations. 
Hence, the unstable periodic orbits of the principal family of EP3 (Lyapunov POs) 
are expected to generate such CS bifurcations as their manifolds extend along the bend degree of freedom. 
Intersections of these manifolds, either self-intersections
or with manifolds from different equilibria, 
generate homoclinic and heteroclinic orbits, respectively \cite{Wiggins_book03}. 
Such orbits can connect remote regions of phase space. 
This phenomenon has been repeatedly underlined in explorations 
of the phase space of a variety of small polyatomic molecules \cite{Farantos09,Mauguiere10,Farantos11}. 
The numerical location of such bi-aymptotic orbits 
is not easy, and this fact makes periodic orbit families even more precious 
in studying the complexity of the molecular phase space at high excitation energies. 

The FR1 POs are associated with a 2:1 resonance region in phase space between 
stretch ($r$) and bend ($\theta$) modes. The CS bifurcation 
generates ``out of nowhere'' stable and unstable PO branches. 
In this way we can 
understand the trapping of (non) reactive trajectories in the roaming mechanism, and
can also assign a DS attached to the NHIM FR1-PO. 

\newpage

\section{Gap time analysis of the roaming region}
\label{gaptime}

In the preceding section, we described the dynamics of roaming reactions. 
We showed that the TTS and the OTS delimit a roaming region inside
which some trajectories may be trapped for long times, and that 
roaming region can be seen as a dynamical complex even if there is no well in the
PES with which this complex can be associated. 
The situation is very similar to that described in Miller's 
unified treatment of statistical theory in the presence of a complex \cite{Miller76a}. 
Our analysis showed that the roaming region can be viewed as a dynamical complex, 
and it is therefore relevant to investigate the validity of the assumption of statistical 
dynamics within the roaming region.

To investigate the nature of dynamics in the roaming region, 
we perform a gap time analysis. We will first briefly 
review the gap time 
approach to reaction rates due to Thiele \cite{Thiele62,Thiele62a}. 
Our exposition follows closely Ref.~\onlinecite{ezra2009microcanonical} to 
which we refer the reader for more information.

Many theoretical 
investigations have been made of the validity of the statistical assumption underlying TST,
focusing on the lifetime and gap time distributions of species involved. 
A non exhaustive list of important work includes the reserarch of Slater \cite{Slater56,Slater59}, 
Bunker \cite{Bunker62,Bunker64}, Bunker and Hase \cite{Bunker73},
Thiele \cite{Thiele62,Thiele62a}, Dumont and Brumer \cite{Dumont86} 
and DeLeon and co-workers \cite{DeLeon81,Berne82}.
Broadly speaking, nonstatistical or non-RRKM 
behavior for a specific unimolecular dissociation reaction 
can arise in two essentially different ways. 
First,  for a specific reaction, non-RRKM behavior can
be observed because reactants are prepared in a specific state 
which violates the assumption of uniform phase space density in the reactant region. 
The second possible origin of non-RRKM behavior 
is due to inherent non-statistical intramolecular dynamics, so-called 
intrinsic non-RRKM behavior \cite{Bunker73}.

\subsection{Gap time approach to unimolecular reaction rates}
\label{gaptime-approach}

\subsubsection{Phase space volumes, gap times and microcanonical RRKM rates}
\label{vol-gaptimes}

To introduce the essential concepts needed, 
we will first treat the case for which the reactant is described by a 
single well to which access is mediated by a single channel (DS).
(For this case, cf.\ Fig.~1 of Ref.~\onlinecite{ezra2009microcanonical}.)
As noted previously, for a given energy, 
a DS constructed from NHIMs divides the energy surface into two distinct
species, reactants and products. 
Furthermore, the DS is composed of two hemispheres, 
one of which intersects trajectories travelling from
reactant to product, and controls exit from the well, 
the other of which intersects trajectories travelling from products to reactants, and
controls the access to the well. 
The hemisphere which controls the access to the well is designated 
$DS_{in}(E)$ and that
which controls the exit from the well $DS_{out}(E)$, 
where the (microcanonical) DS is defined at constant energy $E$.
The distinct phase space regions corresponding to reactants and products are denoted
$\mathcal{M}_{r}$ and $\mathcal{M}_{p}$, respectively. 
The microcanonical density of states for reactant species is:
\begin{equation}
\label{eq:density}
\rho_r(E) = \int_{\mathcal{M}_{r}} d\boldsymbol{x} \; \delta(E-H(\boldsymbol{x})),
\end{equation}
where $\boldsymbol{x} \in \mathbb{R}^{2n}$ designate a phase space point for an $n$ DoF system. 
A similar expression can be written
for the product part of the phase space $\mathcal{M}_{p}$. 
The points on the reactant part of the phase space can be uniquely specified
by coordinates $(\bar{q},\bar{p},\psi)$, 
where $(\bar{q},\bar{p}) \in DS_{in}(E)$ is a point on $DS_{in}(E)$ 
specified by $2(n-1)$ coordinates 
$(\bar{q} , \bar{p})$, and $\psi$ is a time variable. 
The point $\boldsymbol{x}(\bar{q} , \bar{p} , \psi)$ is reached by propagating the initial 
condition $(\bar{q} , \bar{p}) \in DS_{in}(E)$ forward for time $\psi$. As all initial conditions on $DS_{in}(E)$ will leave the reactant region in 
finite time by crossing $DS_{out}(E)$, for each $(\bar{q} , \bar{p}) \in DS_{in}(E)$ 
we can define the \emph{gap time} $s = s(\bar{q} , \bar{p})$, which 
is the time it takes for the incoming trajectory to traverse the reactant region. 
That is, $\boldsymbol{x}(\bar{q} , \bar{p} , \psi = s(\bar{q} , \bar{p})) \in DS_{out}(E)$. 
For the phase point $\boldsymbol{x}(\bar{q},\bar{p},\psi)$, we therefore have $0 \leq \psi \leq s(\bar{q},\bar{p})$.

The coordinate transformation $\boldsymbol{x} \rightarrow (E, \psi, \bar{q}, \bar{p})$ is 
canonical \cite{Thiele62,Binney85,Meyer86} so that 
the phase space volume element is
\begin{equation}
\label{eq:volelemt}
d^{2n}\boldsymbol{x}= dE \; d\psi \; d\sigma,
\end{equation}
with $d\sigma \equiv d^{n-1} \bar{q} \; d^{n-1} \bar{p}$ an element of $2n-2$ dimensional area on the DS. We denote the   
flux across $DS_{in}(E)$ and $DS_{out}(E)$ by $\phi_{in}(E)$ and $\phi_{out}(E)$, respectively, 
and note that $\phi_{in}(E) + \phi_{out}(E) = 0$. 
For our purposes we only need the magnitude of the flux, and so set 
$|\phi_{in}(E)| = |\phi_{out}(E)| \equiv \phi(E)$ 
the magnitude $\phi(E)$ of the flux through dividing surface $DS_{in}(E)$ at energy E is given by
\begin{equation}
\label{eq:flux}
\phi(E)= \left| \int_{DS_{in}(E)} d\sigma \right|,
\end{equation}
where the element of area $d\sigma$ is precisely the restriction to $DS_{in}(E)$ of 
the appropriate flux $(2n-2)$-form, $\omega^{n-1}/(n-1)!$, 
corresponding to the Hamiltonian vector field associated with $H(\boldsymbol{x})$. 
The reactant phase space volume occupied by points 
initiated on the dividing surface $DS_{in}(E)$ with energies between $E$ and $E+dE$ is therefore
\begin{equation}
\label{eq:volume}
dE \int_{DS_{in}(E)} d\sigma \int_0^{s} d\psi = dE \int_{DS_{in}(E)} d\sigma \; s = dE \; \phi(E) \; \bar{s},
\end{equation}
where the \emph{mean gap time} $\bar{s}$ is defined as
\begin{equation}
\label{eq:meangap}
\bar{s} = \frac{1}{\phi(E)} \int_{DS_{in}(E)} d\sigma \; s.
\end{equation}
From this we conclude that the reactant density of state associated with trajectories that enter and exit the well region is
\begin{equation}
\label{eq:reac-density-state}
\rho_r^c(E) = \phi(E) \; \bar{s},
\end{equation}
where the superscript $c$ denotes that this density refers to \textit{crossing} trajectories (some trajectories may be trapped
in the well region and never escape from it). Equation \ref{eq:reac-density-state} is the content of the so-called spectral theorem
\cite{Brumer80,Pollak81,Waalkens05,Waalkens05a,Binney85}.
If all phase space points in the reactant region $\mathcal{M}_{r}$ were to react, we would have $\rho_r^c(E)=\rho_r(E)$, where
$\rho_r(E)$ now denotes the density of states for the full reactant region $\mathcal{M}_{r}$. However, because of the existence of trapped
trajectories, in general we have $\rho_r^c(E) \leq \rho_r(E)$. If $\rho_r^c(E) < \rho_r(E)$ it is then necessary to introduce
corrections to the statistical estimate of the reaction rate \cite{Hase83,Grebenshchikov03,Berblinger94,Berne82,Gray87,Stember07}.

The statistical (RRKM) microcanonical rate for the forward reaction from reactant to products at energy $E$ is given by
\begin{equation}
\label{eq:RRKMrate}
k_{RRKM}(E) = \frac{\phi(E)}{\rho_r(E)},
\end{equation}
and if $\rho_r^c(E)=\rho_r(E)$, we then have
\begin{equation}
\label{eq:RRKMrategaptime}
k_{RRKM}(E) = \frac{1}{\bar{s}}.
\end{equation}
In general the inverse of the mean gap time is given by
\begin{equation}
\label{eq:invmeangap}
\frac{1}{\bar{s}} = \frac{\phi(E)}{\rho_r^c(E)}=k_{RRKM}(E) \left[\frac{\rho_r(E)}{\rho_r^c(E)}\right] \equiv k_{RRKM}^c(E)  \geq k_{RRKM}(E),
\end{equation}
where the superscript in $k_{RRKM}^c(E)$ is for corrected $RRKM$ microcanonical rate.

Generalisation to situations for which the access/exit to the reactant region is controlled by $d$ DSs gives for the corrected RRKM rate
(see Ref.~[\onlinecite{ezra2009microcanonical}])
\begin{equation}
\label{eq:multirate}
k_{RRKM}^c(E) = \frac{\sum_{i=1}^d \phi_i(E)}{\sum_{i=1}^d \bar{s}_{DS_{i,in}(E)} \phi_i (E)}
\end{equation}

\subsubsection{Gap time and lifetime distributions}
\label{distributions}
An important notion in the gap time formulation of TST is the gap time distribution, $P(s;E)$: the probability that a phase space
point on $DS_{in}(E)$ at energy $E$ has a gap time between $s$ and $s+ds$ is equal to $P(s;E)\;ds$. The statistical
assumption of TST is equivalent to the requirement that the gap time distribution is the random, exponential distribution
\begin{equation}
\label{eq:randgapdist}
P(s;E) = k(E) \exp(-k(E)s).
\end{equation}
This distribution is characterised by a single exponential decay constant $k(E)$ function of the energy to which corresponds
the mean gap time $\bar{s}(E)=k(E)^{-1}$.

The lifetime of a phase space point $\boldsymbol{x}(\bar{q},\bar{p},\psi)$ is the time needed for this point to exit the reactant region
$\mathcal{M}_{r}$ by crossing $DS_{out}(E)$ and is then defined as $t=s(\bar{q},\bar{p})-\psi$. It can be shown (see 
Ref.~[\onlinecite{ezra2009microcanonical}]) that the lifetime distribution function $\mathbb{P}(t;E)$ is related to the gap time distribution
by
\begin{equation}
\label{eq:randgapdist2}
\mathbb{P}(t;E) = \frac{1}{\bar{s}} \int_t^{+\infty} ds \; P(s;E).
\end{equation}
An exponential gap time distribution (satisfying the statistical assumption) 
implies that the lifetime distribution is also exponential.

Finally, in addition to the gap time distribution itself, we also consider the integrated gap time distribution $F(t;E)$, which is defined as 
the fraction of trajectories on the DS with gap times $s \geq t$, and is simply the product of the normalized reactant lifetime distribution 
function $\mathbb{P}(t ; E)$ and the mean gap time $\bar{s}$
\begin{equation}
\label{eq:intgapdist}
F(t;E) = \bar{s} \mathbb{P}(t ; E) = \int_t^{+\infty} ds \; P(s;E).
\end{equation}
For the random gap time distribution the integrated gap time distribution is exponential, $F(t;E)=\exp(-kt)$.

\subsection{Trajectory simulations of gap time distributions}
\label{gaptime-traj}

In order to test the statistical assumption for the roaming region 
we analyze the gap time distributions for this 
phase space region. To do so, we sample the OTS  microcanonically 
on the incoming hemisphere 
and integrate  trajectories initiated at these sample points 
until they recross either the OTS or the TTS. 

\begin{figure}[H]
\centering
\includegraphics[scale=0.7]{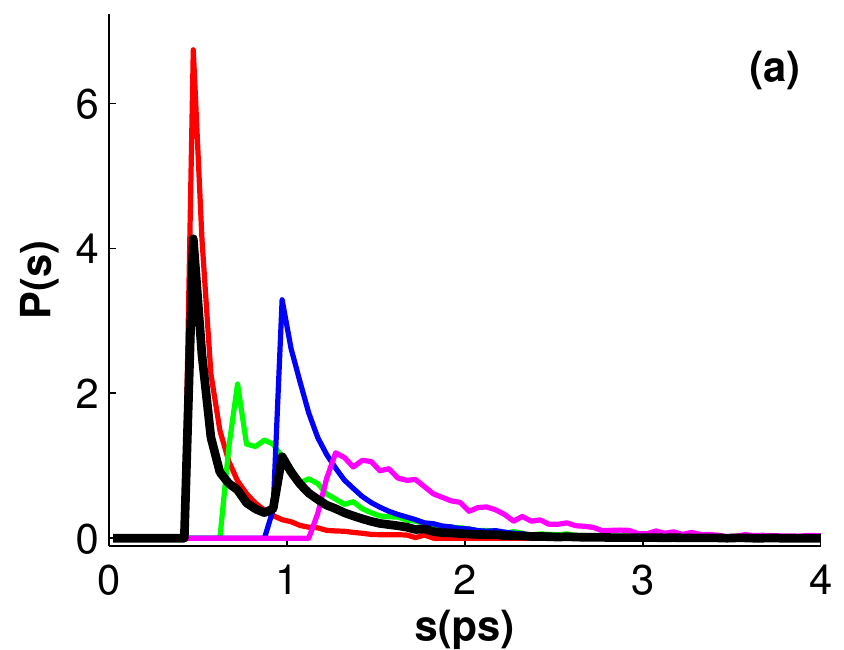}
\includegraphics[scale=0.7]{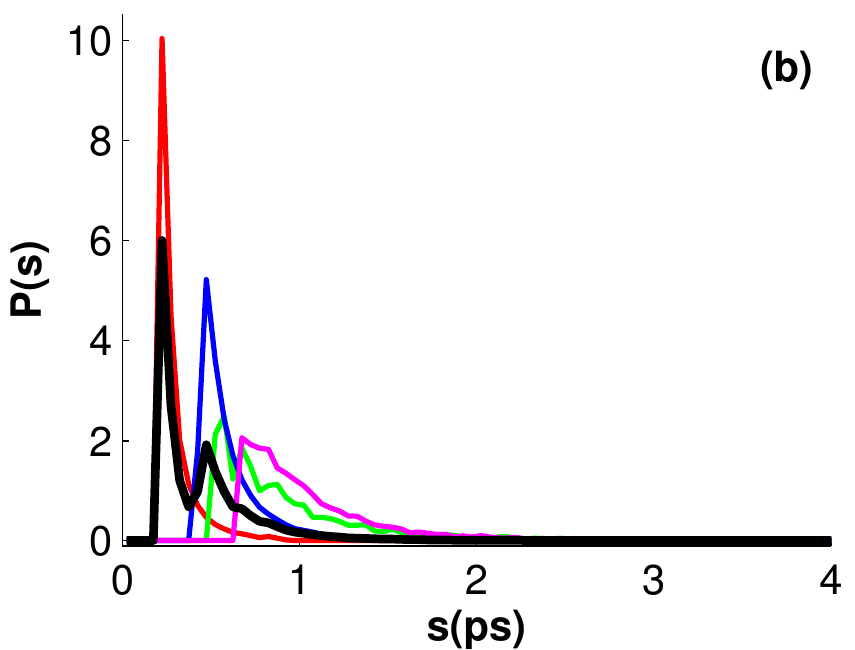}\\
\includegraphics[scale=0.7]{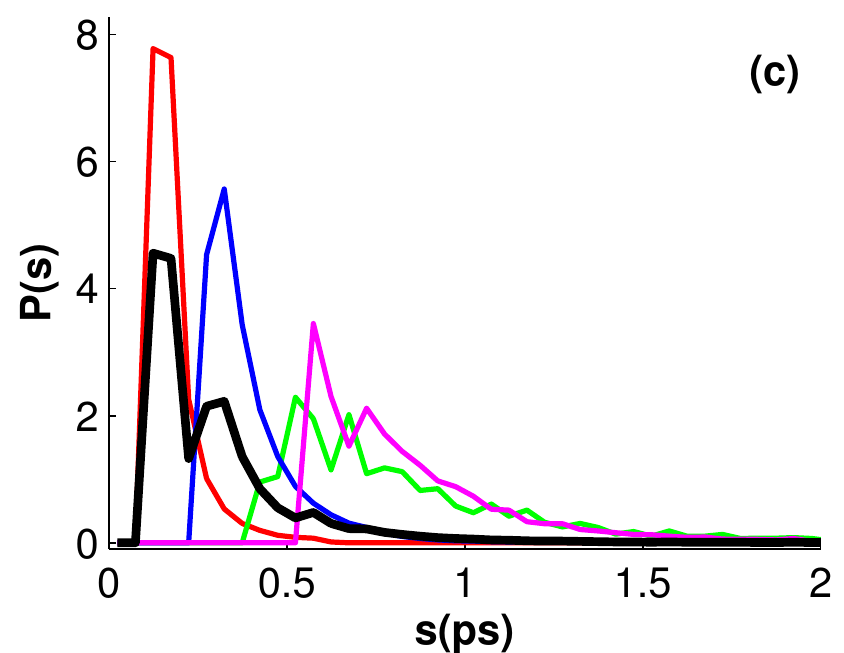}
\includegraphics[scale=0.7]{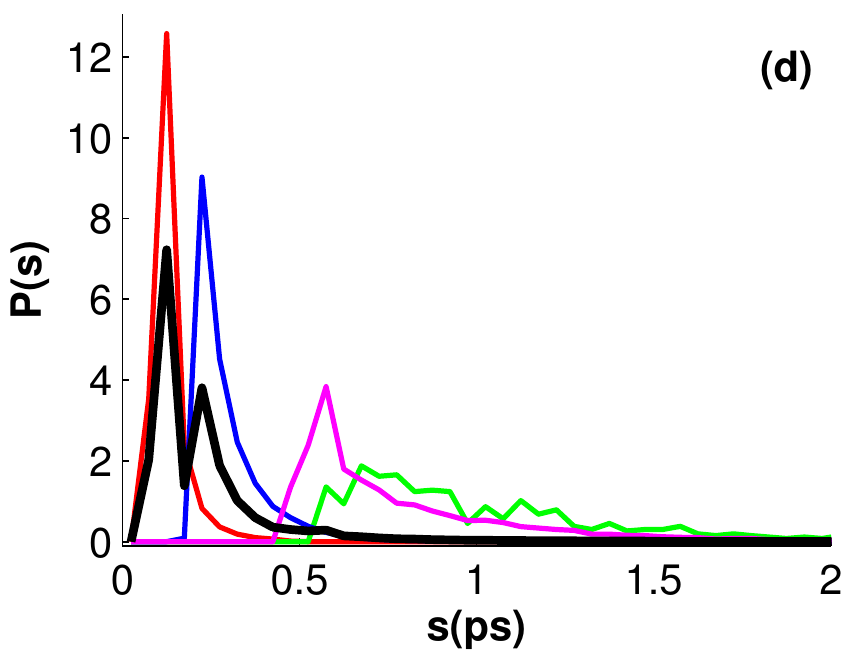}
\caption{Gap time distributions for $\alpha=1$. In each panel, red line denotes the normalised gap time distribution of direct reactive trajectories, green for roaming reactive trajectories, blue for direct non reactive and magenta for roaming non reactive trajectories. 
The thick black curve denotes the normalised gap time distribution for all trajectories.
(a) Energy E=0.5 kcal.mol$^{-1}$. (b) E=1.0 kcal.mol$^{-1}$. 
(c) E=1.5 kcal.mol$^{-1}$. (d) E=2.0 kcal.mol$^{-1}$.}
\label{fig9} 
\end{figure}

\begin{figure}[H]
\centering
\includegraphics[scale=0.7]{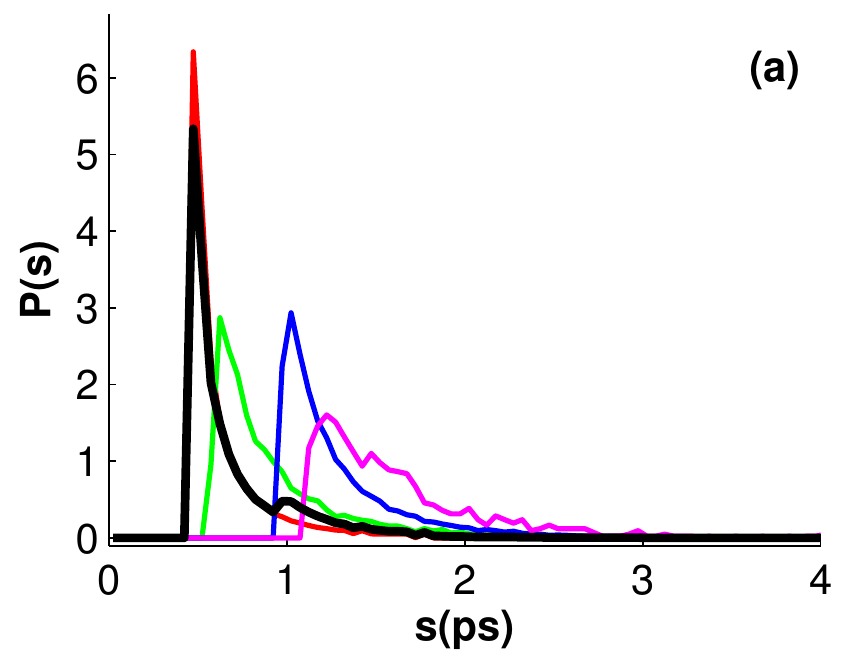}
\includegraphics[scale=0.7]{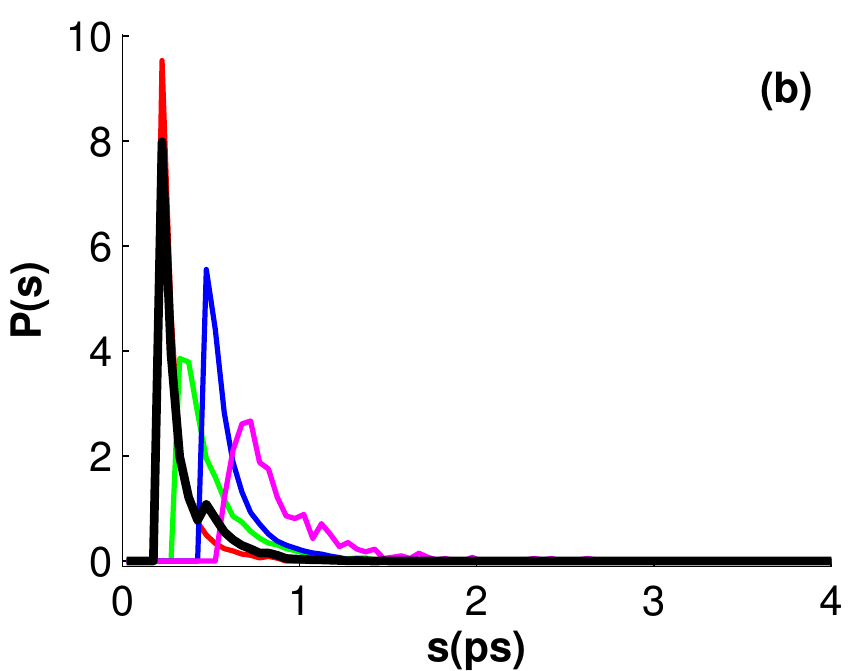}
\includegraphics[scale=0.7]{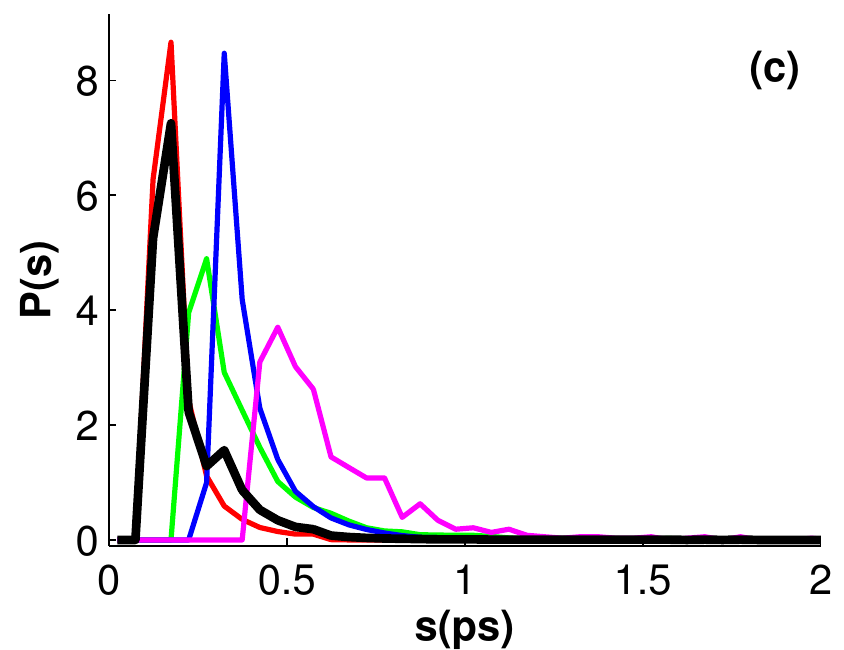}
\includegraphics[scale=0.7]{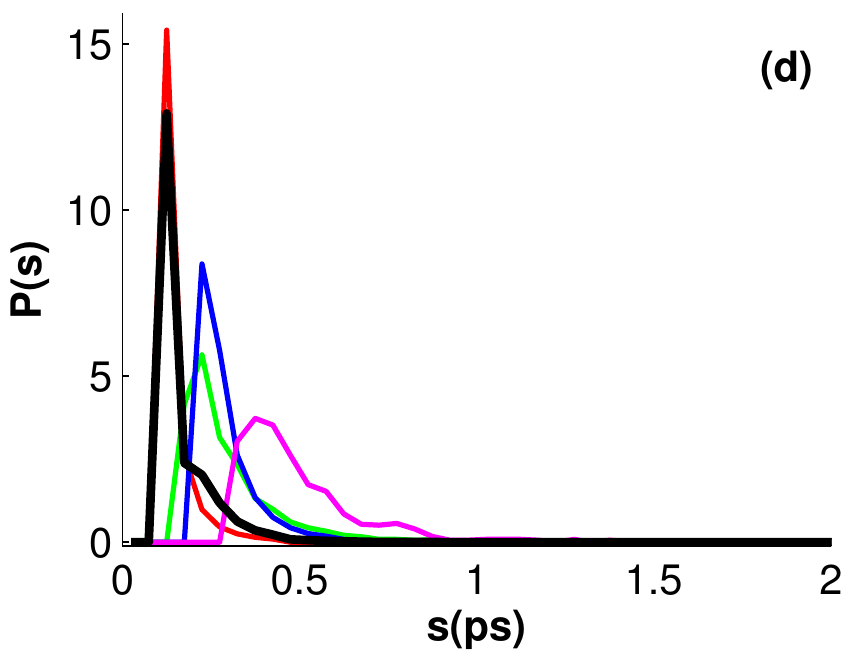}
\caption{Gap time distributions for $\alpha=4$. In each panel, red line denotes the normalised gap time distribution of direct reactive trajectories,
green that for roaming reactive trajectories, blue for direct non reactive trajectories and magenta for roaming non reactive trajectories. 
The thick black curve denote the normalised gap time distribution for all trajectories. 
(a) Energy E=0.5 kcal.mol$^{-1}$.  (b) E=1.0 kcal.mol$^{-1}$.  
(c) E=1.5 kcal.mol$^{-1}$. (d) E=2.0 kcal.mol$^{-1}$.}
\label{fig10} 
\end{figure}

Gap time
distributions obtained from these simulations are shown in Figs.~\ref{fig9} and \ref{fig10} for 
$\alpha=1$ and $\alpha=4$, respectively. Each of these figures has 4 panels, corresponding to different
energies. In each panel we show the
normalised gap time distributions for each of the four classes of 
trajectories we defined earlier, as well as the
gap time distribution for all the trajectories taken together. 
Details are given in the captions of the figures.

\begin{figure}[H]
\centering
\includegraphics[scale=0.7]{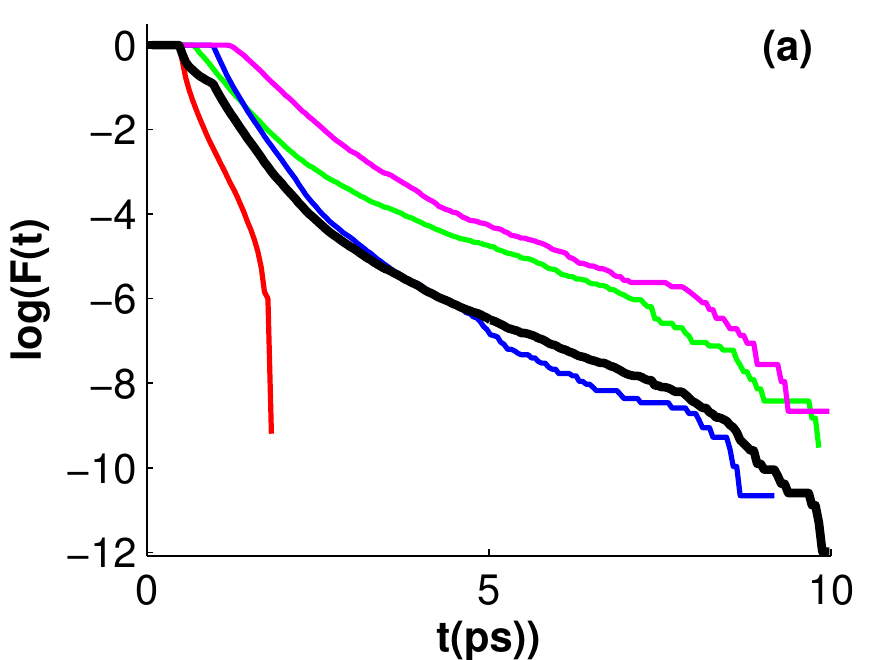}
\includegraphics[scale=0.7]{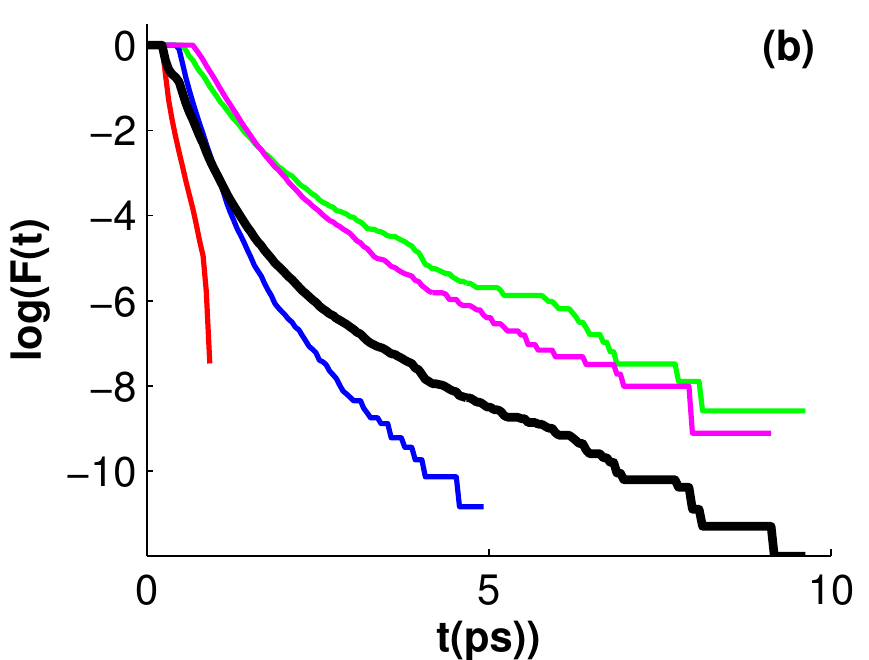}
\includegraphics[scale=0.7]{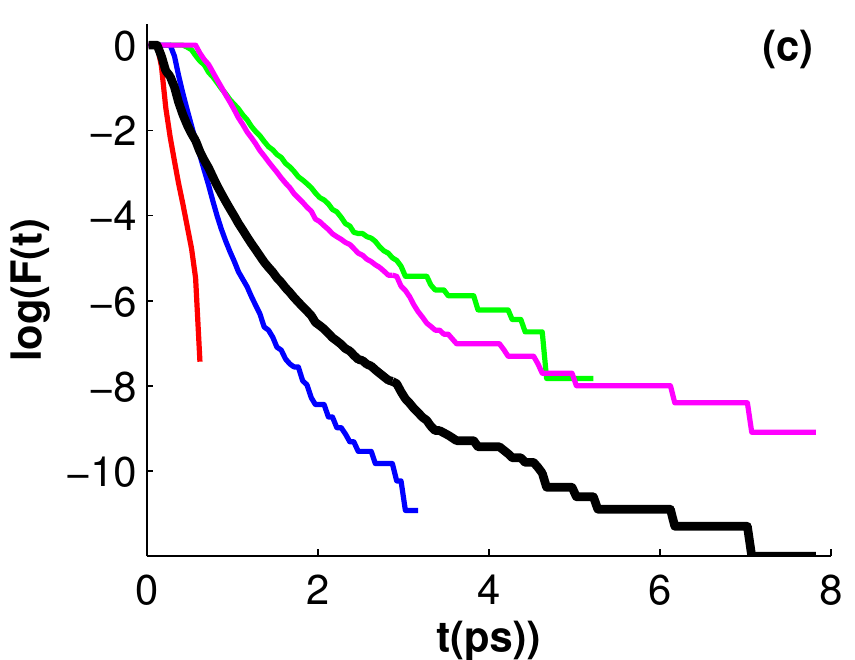}
\includegraphics[scale=0.7]{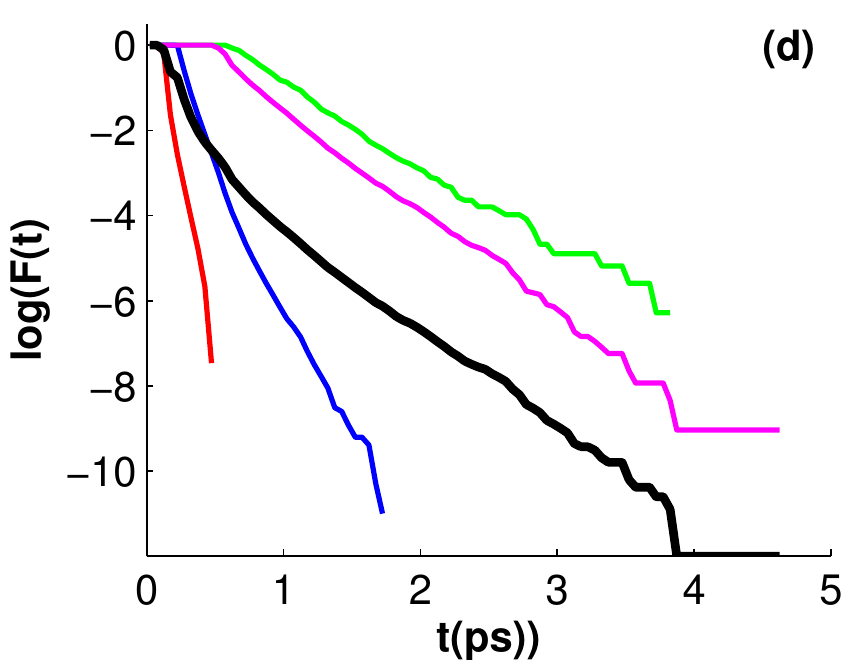}
\caption{The logarithm of the lifetime distributions for $\alpha=1$. In each panel, red line denotes the normalised 
logarithm of the lifetime distribution of direct reactive trajectories, green for roaming reactive trajectories, blue for direct non reactive 
and magenta for roaming non reactive trajectories. The thick black curve denotes the normalised logarithm  of the lifetime for all trajectories. 
(a) Energy E=0.5 kcal.mol$^{-1}$. (b) E=1.0 kcal.mol$^{-1}$. (c) E=1.5 kcal.mol$^{-1}$. (d) E=2.0 kcal.mol$^{-1}$.}
\label{fig11} 
\end{figure}

\begin{figure}[H]
\centering
\includegraphics[scale=0.7]{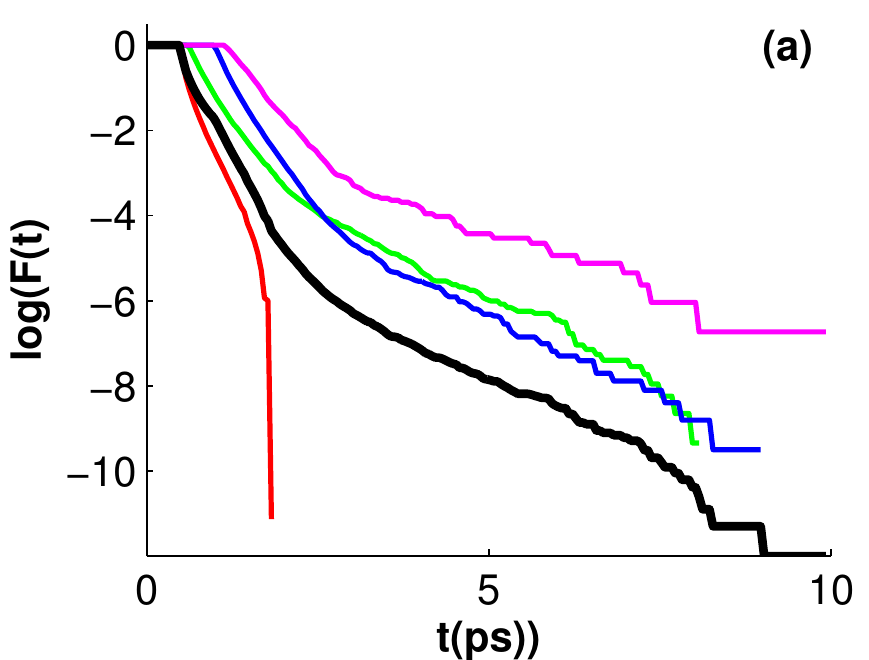}
\includegraphics[scale=0.7]{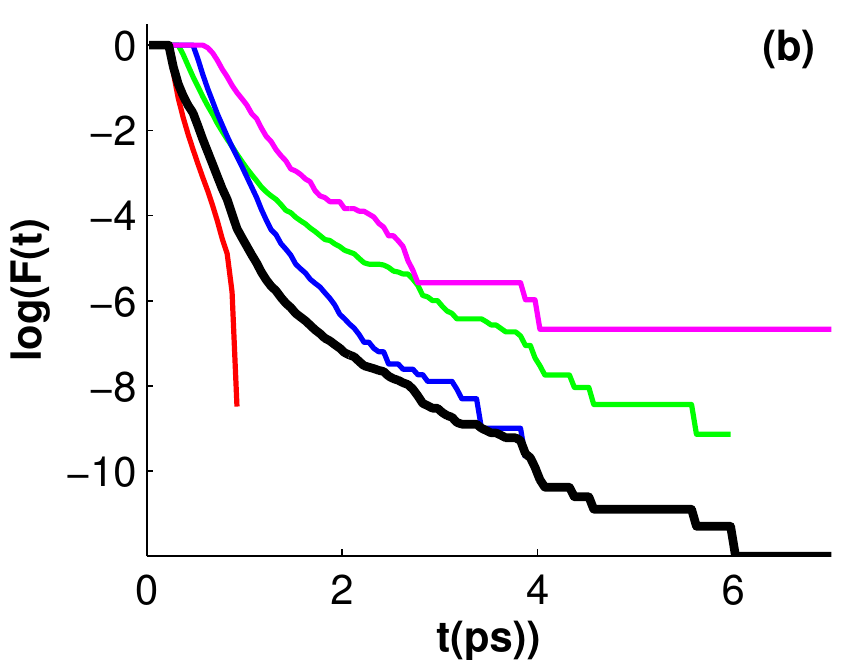}
\includegraphics[scale=0.7]{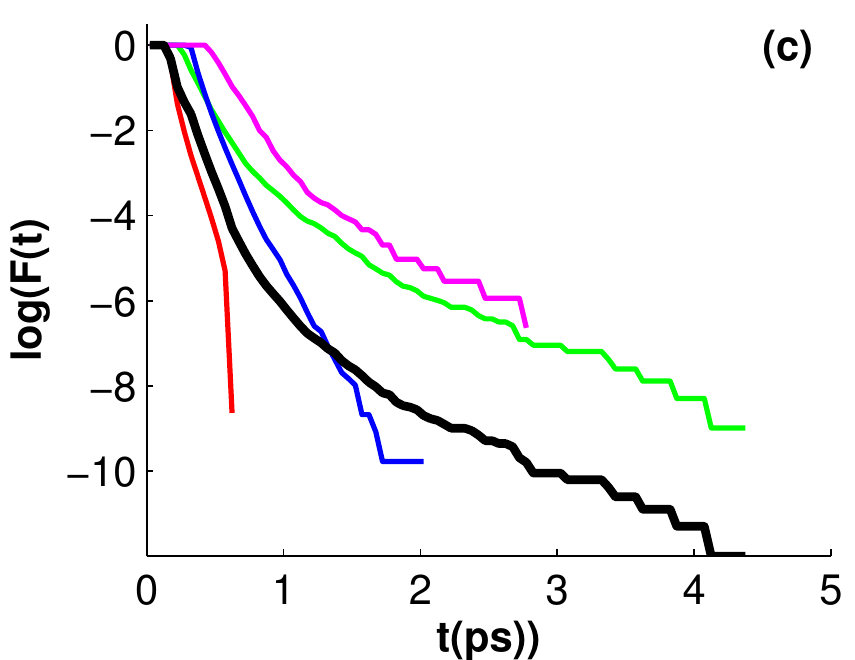}
\includegraphics[scale=0.7]{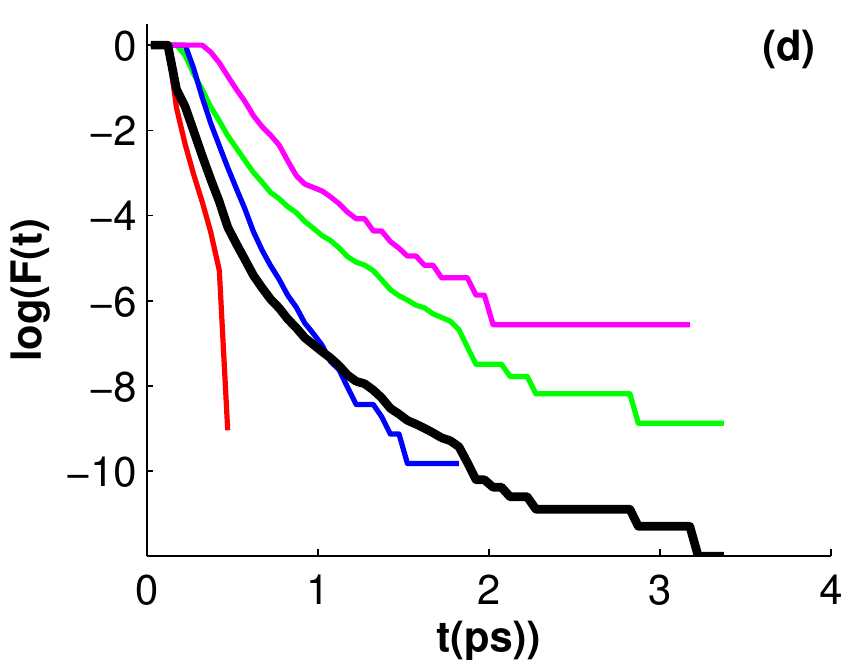}
\caption{The logarithm of the lifetime distributions for $\alpha=4$. 
In each panel, red line denotes the normalised logarithm  of the lifetime
distribution of direct reactive trajectories, green for roaming reactive trajectories, blue for direct non reactive  and 
magenta for roaming non reactive trajectories. The thick black curve denotes the normalised logarithm of the lifetime for all trajectories.
(a) Energy E=0.5 kcal.mol$^{-1}$. (b) E=1.0 kcal.mol$^{-1}$. (c) E=1.5 kcal.mol$^{-1}$. (d) E=2.0 kcal.mol$^{-1}$.}
\label{fig12} 
\end{figure}

The integrated gap time distributions for the same samples used in Figs~\ref{fig9} and \ref{fig10} 
are shown in Figs~\ref{fig11} and \ref{fig12}, respectively. 
As seen in these figures, the gap time distributions, as well as the integrated gap time
distributions, exhibit significant deviation from random (exponential) distributions, 
indicating that
the statistical dynamical assumption of TST is not satisfied for motion in the roaming region.

\begin{figure}[H]
\centering
\includegraphics[scale=0.6]{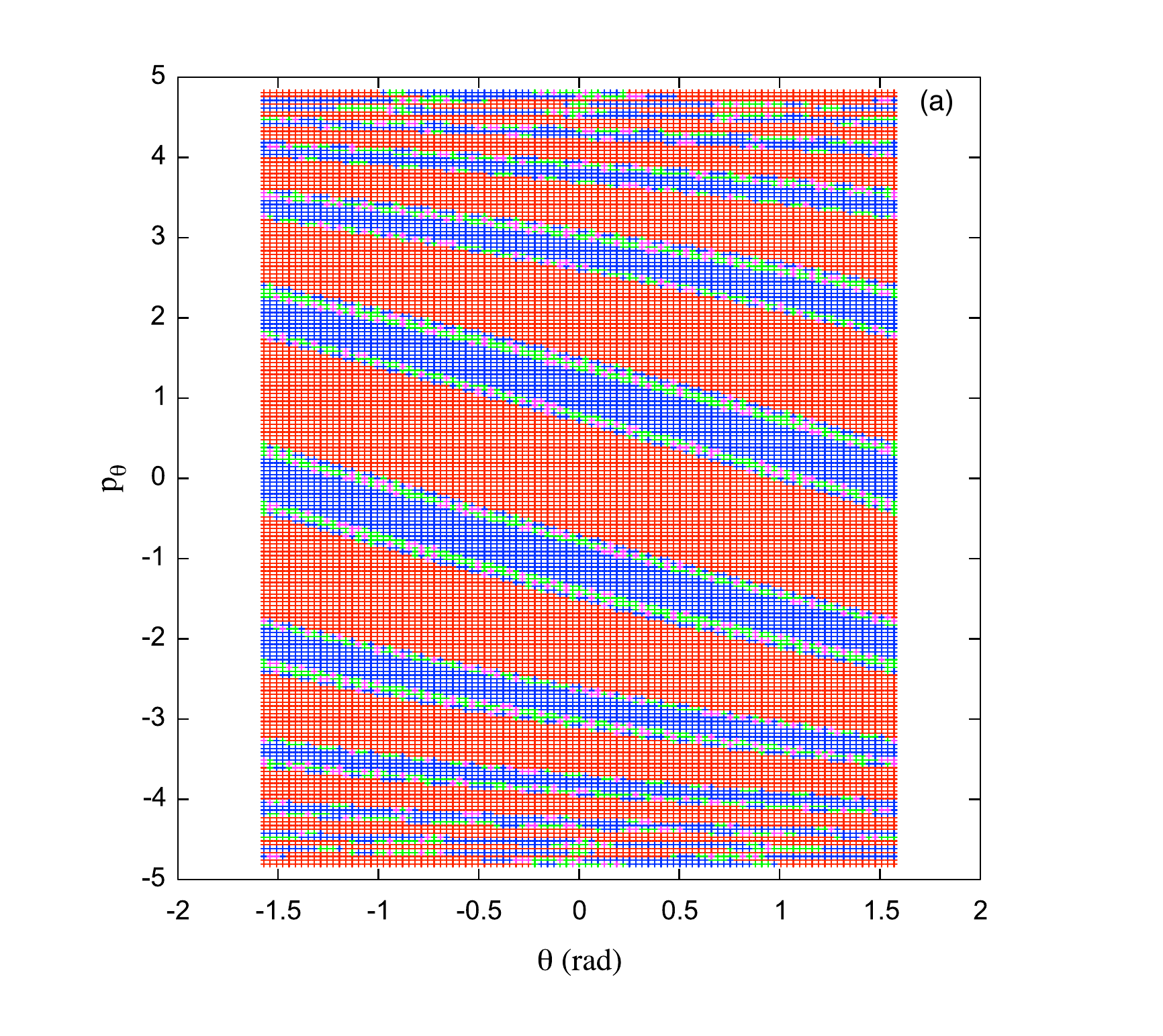}
\includegraphics[scale=0.6]{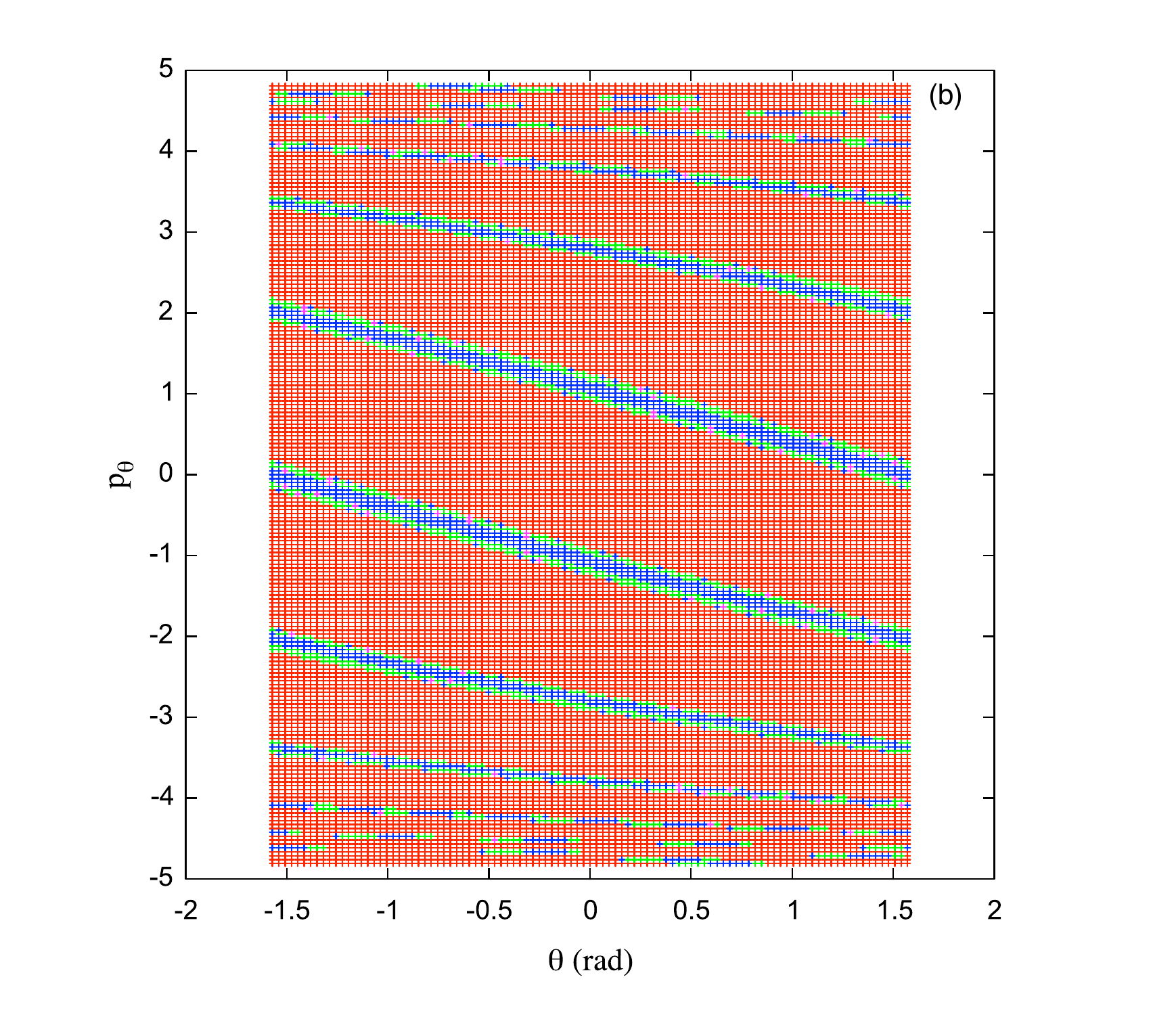}
\caption{Distribution of the different types of trajectories on the OTS. 
(a) $\alpha=1$, energy E=0.5 kcal.mol$^{-1}$.
(b) $\alpha=4$, energy E=0.5 kcal.mol$^{-1}$}
\label{fig13} 
\end{figure}

\begin{figure}[H]
\centering
\includegraphics[scale=1.5]{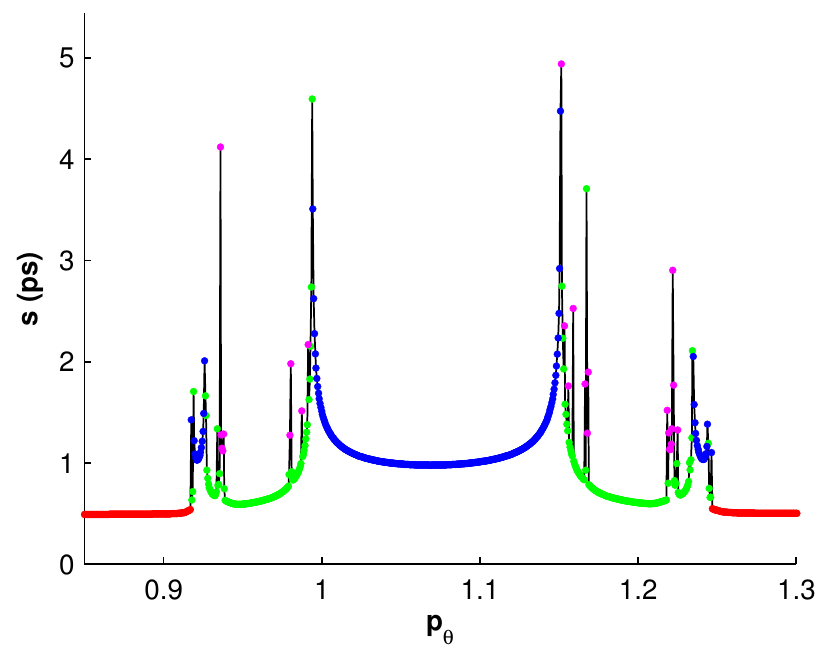}
\caption{The fractal nature of the boundaries between different types of trajectories on the DS. 
Initial points are selected along the line $\theta=0$ on the OTS, with $\alpha=4$ and energy E=0.5 kcal.mol$^{-1}$. 
The vertical axis shows the gap time and sampling points are assigned a colour to mark their type. 
Red: direct reactive trajectories, green: roaming reactive trajectories, blue: direct non reactive and 
magenta: roaming non reactive trajectories.}
\label{fig14} 
\end{figure}

In Fig~\ref{fig13} we plot the sampled trajectory initial conditions on the OTS 
in the $(\theta, p_\theta)$ plane; different colors are used to
represent trajectory intial conditions belonging to different classes. 
This plot reveals a succession of \emph{bands} of
different types on the DS (see, for example, ref.\ \onlinecite{Nagahata13} and references therein). 
The arrangement of bands can be very complicated (fractal) \cite{Grice87}.
In Fig~\ref{fig14} we plot gap time versus $p_\theta$ for
initial conditions on the OTS at fixed $\theta=0$ for a range of $p_\theta$ values. 
The plot shows the fractal nature of bands associated with different 
trajectory types, and indicates that gap times diverge
at the boundary between bands associated with two different trajectory types \cite{Pechukas77,Mauguiere13}.
An infinitely fine sampling of the $p_\theta$ axis would presumably reveal
a set of measure zero of initial conditions 
for which the gap times are infinite. Infinite
gap times correspond to trajectories trapped forever in the roaming regions, 
and such trajectories are on the stable  invariant manifolds
of stationary objects in the roaming region, such as the FR1 PO and its period doubling bifurcations.

\newpage

\section{Summary and Conclusion}
\label{sec:conclusion}

The model Hamiltonian for the reaction CH$_4^+ \rightarrow$  CH$_3^+$ + H 
proposed by Chesnavich \cite{Chesnavich1986} to study transition state switching  in ion-molecule
reactions has been employed to investigate roaming dynamics. 
The Chesnavich model supports multiple transition states and, 
despite its simplicity, is endowed with all the essential characteristics of systems previously found 
to exhibit the roaming mechanism.

Using concepts and methods from non-linear mechanics, early/late or tight/loose transition 
states are identified with time invariant objects in phase space, which are
dividing surfaces in phase space associated with NHIMs -- normally hyperbolic invariant manifolds.
For two degree of freedom systems NHIMs are unstable periodic orbits 
which define the boundaries of locally non-recrossing dividing surfaces 
assumed in statistical reaction rate theories
such as TST.  The roaming region of phase space 
is itself unambiguously defined by these 
dividing surfaces.

By constructing continuation/bifurcation diagrams of periodic orbits for 
two values of the parameter in the Chesnavich Hamiltonian model 
controlling the early versus late nature of the transition state, and 
using the total energy as a second parameter, 
we identify phase space regions associated with roaming reaction pathways
(i.e., trapping in the roaming region). 
The classical dynamics of the system are investigated 
by microcanonically 
sampling the outer OTS DS and assigning trajectories to four different classes:
direct reactive and direct non-reactive,
which describe the formation of molecular and radical products respectively, and roaming reactive and 
roaming non reactive, which folow alternative pathways to formation of molecular and radical products.

We identify the TTS and OTS with dividing surfaces associated with unstable periodic orbits
of the TTS-PO and OTS-PO families. 
Additional PO families such as the FR1 POs reveal alternative reaction pathways, 
the roaming pathway, and define  region in phase space 
associated with a 2:1 resonance between the
stretch ($r$) and the bend ($\theta$) modes. 
The CS bifurcations 
generate ``out of nowhere'' a branch with stable and a branch with unstable periodic orbits. 
In this way we can 
understand the dynamical origin of the 
trapping of (non) reactive trajectories in the roaming region. 

To investigate the validity of the
assumption of statistical dynamics for the microcanonical ensembles we consider, 
we have analysed gap time distributions at several energies.
Lifetime distributions exhibit multiple exponential dissociation rates at long times,
violating the assumption of random gap times underlying statistical theory.

By plotting the outcome for trajectory initial conditions 
initiated on the $(\theta, p_\theta)$ at the OTS DS, we
observe a regular succession of `bands' of different types of trajectories. 
Our numerical results indicate the existence of a fractal band structure, where
the gap time diverges at the boundary between distinct trajectory types.
Such divergent gap times are associated with initial conditions on 
the stable manifold
of invariant objects in the roaming region such as the FR1 PO and its period doubling bifurcations.

It is worth emphasizing that the concepts, theory and algorithms described 
here for two degrees of freedom systems can in principle be straightforwardly extended 
to higher dimensional systems. Nevertheless, substantial technical difficulties 
need to be overcome for accurate computation of NHIM-DS 
for higher dimensional systems. 

\section*{Acknowledgments}

This work is supported by the National Science Foundation under Grant No.\ CHE-1223754 (to GSE).
FM, PC, and SW  acknowledge the support of the  Office of Naval Research (Grant No.~N00014-01-1-0769),
the Leverhulme Trust, and the Engineering and Physical Sciences Research Council (Grant No.~ EP/K000489/1).

\newpage


\begin{thebibliography}{10}
\expandafter\ifx\csname bibnamefont\endcsname\relax
  \def\bibnamefont#1{#1}\fi
\expandafter\ifx\csname bibfnamefont\endcsname\relax
  \def\bibfnamefont#1{#1}\fi
\expandafter\ifx\csname url\endcsname\relax
  \def\url#1{\texttt{#1}}\fi
\expandafter\ifx\csname urlprefix\endcsname\relax\def\urlprefix{URL }\fi
\expandafter\ifx\csname bibinfo\endcsname\relax \def\bibinfo#1#2{#2}\fi
\expandafter\ifx\csname eprint\endcsname\relax \def\eprint#1{#1}\fi

\bibitem{Ashfold06}
\bibinfo{author}{\bibfnamefont{M.~N.~R.} \bibnamefont{Ashfold}},
  \bibinfo{author}{\bibfnamefont{N.}~\bibnamefont{Nahler}},
  \bibinfo{author}{\bibfnamefont{A.}~\bibnamefont{Orr-Ewing}},
  \bibinfo{author}{\bibfnamefont{O.}~\bibnamefont{Vieuxmaire}},
  \bibinfo{author}{\bibfnamefont{R.}~\bibnamefont{Toomes}},
  \bibinfo{author}{\bibfnamefont{T.}~\bibnamefont{Kitsopoulos}},
  \bibinfo{author}{\bibfnamefont{I.}~\bibnamefont{Garcia}},
  \bibinfo{author}{\bibfnamefont{D.}~\bibnamefont{Chestakov}},
  \bibinfo{author}{\bibfnamefont{S.}~\bibnamefont{Wu}}, \bibnamefont{and}
  \bibinfo{author}{\bibfnamefont{D.}~\bibnamefont{Parker}},
  \bibinfo{journal}{Phys. Chem. Chem. Phys.} \textbf{\bibinfo{volume}{8}},
  \bibinfo{pages}{26} (\bibinfo{year}{2006}).

\bibitem{Mukamel2000}
\bibinfo{author}{\bibfnamefont{S.}~\bibnamefont{Mukamel}},
  \bibinfo{journal}{Ann. Rev. Phys. Chem.} \textbf{\bibinfo{volume}{51}},
  \bibinfo{pages}{691} (\bibinfo{year}{2000}).

\bibitem{Forst03}
\bibinfo{author}{\bibfnamefont{W.}~\bibnamefont{Forst}},
  \emph{\bibinfo{title}{{Unimolecular Reactions}}}
  (\bibinfo{publisher}{Cambridge University Press},
  \bibinfo{address}{Cambridge}, \bibinfo{year}{2003}).

\bibitem{Baer96}
\bibinfo{author}{\bibfnamefont{T.}~\bibnamefont{Baer}} \bibnamefont{and}
  \bibinfo{author}{\bibfnamefont{W.~L.} \bibnamefont{Hase}},
  \emph{\bibinfo{title}{{Unimolecular Reaction Dynamics}}}
  (\bibinfo{publisher}{Oxford University Press}, \bibinfo{address}{New York},
  \bibinfo{year}{1996}).

\bibitem{Levine09}
\bibinfo{author}{\bibfnamefont{R.~D.} \bibnamefont{Levine}},
  \emph{\bibinfo{title}{{Molecular Reaction Dynamics}}}
  (\bibinfo{publisher}{Cambridge University Press}, \bibinfo{year}{2009}).

\bibitem{townsend2004roaming}
\bibinfo{author}{\bibfnamefont{D.}~\bibnamefont{Townsend}},
  \bibinfo{author}{\bibfnamefont{S.~A.} \bibnamefont{Lahankar}},
  \bibinfo{author}{\bibfnamefont{S.~K.} \bibnamefont{Lee}},
  \bibinfo{author}{\bibfnamefont{S.~D.} \bibnamefont{Chambreau}},
  \bibinfo{author}{\bibfnamefont{A.~G.} \bibnamefont{Suits}},
  \bibinfo{author}{\bibfnamefont{X.}~\bibnamefont{Zhang}},
  \bibinfo{author}{\bibfnamefont{J.}~\bibnamefont{Rheinecker}},
  \bibinfo{author}{\bibfnamefont{L.~B.} \bibnamefont{Harding}},
  \bibnamefont{and} \bibinfo{author}{\bibfnamefont{J.~M.}
  \bibnamefont{Bowman}}, \bibinfo{journal}{Science}
  \textbf{\bibinfo{volume}{306}}(\bibinfo{number}{5699}), \bibinfo{pages}{1158}
  (\bibinfo{year}{2004}).

\bibitem{zee:1664}
\bibinfo{author}{\bibfnamefont{R.~D.} \bibnamefont{van Zee}},
  \bibinfo{author}{\bibfnamefont{M.~F.} \bibnamefont{Foltz}}, \bibnamefont{and}
  \bibinfo{author}{\bibfnamefont{C.~B.} \bibnamefont{Moore}},
  \bibinfo{journal}{J. Chem. Phys.}
  \textbf{\bibinfo{volume}{99}}(\bibinfo{number}{3}), \bibinfo{pages}{1664}
  (\bibinfo{year}{1993}),
  \urlprefix\url{http://link.aip.org/link/?JCP/99/1664/1}.

\bibitem{Yu11}
\bibinfo{author}{\bibfnamefont{H.~G.} \bibnamefont{Yu}},
  \bibinfo{journal}{Phys. Scr.} \textbf{\bibinfo{volume}{84}},
  \bibinfo{pages}{028104} (\bibinfo{year}{2011}).

\bibitem{bowman2011roaming}
\bibinfo{author}{\bibfnamefont{J.~M.} \bibnamefont{Bowman}} \bibnamefont{and}
  \bibinfo{author}{\bibfnamefont{B.~C.} \bibnamefont{Shepler}},
  \bibinfo{journal}{Ann. Rev. Phys. Chem.} \textbf{\bibinfo{volume}{62}},
  \bibinfo{pages}{531} (\bibinfo{year}{2011}).

\bibitem{Heidrich95}
\bibinfo{editor}{\bibfnamefont{D.}~\bibnamefont{Heidrich}}, ed.,
  \emph{\bibinfo{title}{{The Reaction Path in Chemistry: Current Approaches and
  Perspectives}}} (\bibinfo{publisher}{Springer}, \bibinfo{address}{New York},
  \bibinfo{year}{1995}).

\bibitem{Sun02}
\bibinfo{author}{\bibfnamefont{L.~P.} \bibnamefont{Sun}},
  \bibinfo{author}{\bibfnamefont{K.~Y.} \bibnamefont{Song}}, \bibnamefont{and}
  \bibinfo{author}{\bibfnamefont{W.~L.} \bibnamefont{Hase}},
  \bibinfo{journal}{Science} \textbf{\bibinfo{volume}{296}},
  \bibinfo{pages}{875} (\bibinfo{year}{2002}).

\bibitem{Lopez07}
\bibinfo{author}{\bibfnamefont{J.~G.} \bibnamefont{Lopez}},
  \bibinfo{author}{\bibfnamefont{G.}~\bibnamefont{Vayner}},
  \bibinfo{author}{\bibfnamefont{U.}~\bibnamefont{Lourderaj}},
  \bibinfo{author}{\bibfnamefont{S.~V.} \bibnamefont{Addepalli}},
  \bibinfo{author}{\bibfnamefont{S.}~\bibnamefont{Kato}},
  \bibinfo{author}{\bibfnamefont{W.~A.} \bibnamefont{Dejong}},
  \bibinfo{author}{\bibfnamefont{T.~L.} \bibnamefont{Windus}},
  \bibnamefont{and} \bibinfo{author}{\bibfnamefont{W.~L.} \bibnamefont{Hase}},
  \bibinfo{journal}{J. Am. Chem. Soc.} \textbf{\bibinfo{volume}{129}},
  \bibinfo{pages}{9976} (\bibinfo{year}{2007}).

\bibitem{Mikosch08}
\bibinfo{author}{\bibfnamefont{J.}~\bibnamefont{Mikosch}},
  \bibinfo{author}{\bibfnamefont{S.}~\bibnamefont{Trippel}},
  \bibinfo{author}{\bibfnamefont{C.}~\bibnamefont{Eichhorn}},
  \bibinfo{author}{\bibfnamefont{R.}~\bibnamefont{Otto}},
  \bibinfo{author}{\bibfnamefont{U.}~\bibnamefont{Lourderaj}},
  \bibinfo{author}{\bibfnamefont{J.~X.} \bibnamefont{Zhang}},
  \bibinfo{author}{\bibfnamefont{W.~L.} \bibnamefont{Hase}},
  \bibinfo{author}{\bibfnamefont{M.}~\bibnamefont{Weidemuller}},
  \bibnamefont{and} \bibinfo{author}{\bibfnamefont{R.}~\bibnamefont{Wester}},
  \bibinfo{journal}{Science} \textbf{\bibinfo{volume}{319}},
  \bibinfo{pages}{183} (\bibinfo{year}{2008}).

\bibitem{Zhang10}
\bibinfo{author}{\bibfnamefont{J.}~\bibnamefont{Zhang}},
  \bibinfo{author}{\bibfnamefont{J.}~\bibnamefont{Mikosch}},
  \bibinfo{author}{\bibfnamefont{S.}~\bibnamefont{Trippel}},
  \bibinfo{author}{\bibfnamefont{R.}~\bibnamefont{Otto}},
  \bibinfo{author}{\bibfnamefont{M.}~\bibnamefont{{Weidem\"uller}}},
  \bibinfo{author}{\bibfnamefont{R.}~\bibnamefont{Wester}}, \bibnamefont{and}
  \bibinfo{author}{\bibfnamefont{W.~L.} \bibnamefont{Hase}},
  \bibinfo{journal}{J. Phys. Chem. Lett.}
  \textbf{\bibinfo{volume}{1}}(\bibinfo{number}{18}), \bibinfo{pages}{2747}
  (\bibinfo{year}{2010}).

\bibitem{NorthScience2012}
\bibinfo{author}{\bibfnamefont{M.~P.} \bibnamefont{Grubb}},
  \bibinfo{author}{\bibfnamefont{M.~L.} \bibnamefont{Warter}},
  \bibinfo{author}{\bibfnamefont{H.}~\bibnamefont{Xiao}},
  \bibinfo{author}{\bibfnamefont{S.}~\bibnamefont{Maeda}},
  \bibinfo{author}{\bibfnamefont{K.}~\bibnamefont{Morokuma}}, \bibnamefont{and}
  \bibinfo{author}{\bibfnamefont{S.~W.} \bibnamefont{North}},
  \bibinfo{journal}{Science} \textbf{\bibinfo{volume}{335}},
  \bibinfo{pages}{1075} (\bibinfo{year}{2012}).

\bibitem{Ulusoy13}
\bibinfo{author}{\bibfnamefont{I.~G.} \bibnamefont{Ulusoy}},
  \bibinfo{author}{\bibfnamefont{J.~F.} \bibnamefont{Stanton}},
  \bibnamefont{and}
  \bibinfo{author}{\bibfnamefont{R.}~\bibnamefont{Hernandez}},
  \bibinfo{journal}{J. Phys. Chem. A} \textbf{\bibinfo{volume}{117}},
  \bibinfo{pages}{7553} (\bibinfo{year}{2013}).

\bibitem{Ulusoy13b}
\bibinfo{author}{\bibfnamefont{I.~G.} \bibnamefont{Ulusoy}},
  \bibinfo{author}{\bibfnamefont{J.~F.} \bibnamefont{Stanton}},
  \bibnamefont{and}
  \bibinfo{author}{\bibfnamefont{R.}~\bibnamefont{Hernandez}},
  \bibinfo{journal}{J. Phys. Chem. A} \textbf{\bibinfo{volume}{117}},
  \bibinfo{pages}{10567} (\bibinfo{year}{2013}).

\bibitem{Truhlar1984}
\bibinfo{author}{\bibfnamefont{D.~G.} \bibnamefont{Truhlar}} \bibnamefont{and}
  \bibinfo{author}{\bibfnamefont{B.~C.} \bibnamefont{Garrett}},
  \bibinfo{journal}{Ann. Rev. Phys. Chem.} \textbf{\bibinfo{volume}{35}},
  \bibinfo{pages}{159} (\bibinfo{year}{1984}).

\bibitem{Carpenter84}
\bibinfo{author}{\bibfnamefont{B.~K.} \bibnamefont{Carpenter}},
  \emph{\bibinfo{title}{{Determination of Organic Reaction Mechanisms}}}
  (\bibinfo{publisher}{Wiley}, \bibinfo{address}{New York},
  \bibinfo{year}{1984}).

\bibitem{Wales03}
\bibinfo{author}{\bibfnamefont{D.~J.} \bibnamefont{Wales}},
  \emph{\bibinfo{title}{{Energy Landscapes}}} (\bibinfo{publisher}{Cambridge
  University Press}, \bibinfo{address}{Cambridge}, \bibinfo{year}{2003}).

\bibitem{shepler2011roaming}
\bibinfo{author}{\bibfnamefont{B.~C.} \bibnamefont{Shepler}},
  \bibinfo{author}{\bibfnamefont{Y.}~\bibnamefont{Han}}, \bibnamefont{and}
  \bibinfo{author}{\bibfnamefont{J.~M.} \bibnamefont{Bowman}},
  \bibinfo{journal}{J. Phys. Chem. Lett.}
  \textbf{\bibinfo{volume}{2}}(\bibinfo{number}{7}), \bibinfo{pages}{834}
  (\bibinfo{year}{2011}).

\bibitem{Harding_et_al_2012}
\bibinfo{author}{\bibfnamefont{L.~B.} \bibnamefont{Harding}},
  \bibinfo{author}{\bibfnamefont{S.~J.} \bibnamefont{Klippenstein}},
  \bibnamefont{and} \bibinfo{author}{\bibfnamefont{A.~W.}
  \bibnamefont{Jasper}}, \bibinfo{journal}{J. Phys. Chem. A}
  \textbf{\bibinfo{volume}{116}}, \bibinfo{pages}{6967} (\bibinfo{year}{2012}).

\bibitem{Wigner38}
\bibinfo{author}{\bibfnamefont{E.~P.} \bibnamefont{Wigner}},
  \bibinfo{journal}{Trans. Faraday Soc.} \textbf{\bibinfo{volume}{34}},
  \bibinfo{pages}{29} (\bibinfo{year}{1938}).

\bibitem{Wiggins08}
\bibinfo{author}{\bibfnamefont{H.}~\bibnamefont{Waalkens}},
  \bibinfo{author}{\bibfnamefont{R.}~\bibnamefont{Schubert}}, \bibnamefont{and}
  \bibinfo{author}{\bibfnamefont{S.}~\bibnamefont{Wiggins}},
  \bibinfo{journal}{Nonlinearity} \textbf{\bibinfo{volume}{21}},
  \bibinfo{pages}{R1} (\bibinfo{year}{2008}).

\bibitem{Wiggins_book1994}
\bibinfo{author}{\bibfnamefont{S.}~\bibnamefont{Wiggins}},
  \emph{\bibinfo{title}{{Normally hyperbolic invariant manifolds in dynamical
  systems}}} (\bibinfo{publisher}{Springer-Verlag, New York},
  \bibinfo{year}{1994}).

\bibitem{ezra2009microcanonical}
\bibinfo{author}{\bibfnamefont{G.~S.} \bibnamefont{Ezra}},
  \bibinfo{author}{\bibfnamefont{H.}~\bibnamefont{Waalkens}}, \bibnamefont{and}
  \bibinfo{author}{\bibfnamefont{S.}~\bibnamefont{Wiggins}},
  \bibinfo{journal}{J. Chem. Phys.} \textbf{\bibinfo{volume}{130}},
  \bibinfo{pages}{164118} (\bibinfo{year}{2009}).

\bibitem{Wiggins_book03}
\bibinfo{author}{\bibfnamefont{S.}~\bibnamefont{Wiggins}},
  \emph{\bibinfo{title}{{Introduction to Applied Nonlinear Dynamical Systems
  and Chaos}}} (\bibinfo{publisher}{Springer-Verlag, New York},
  \bibinfo{year}{2003}), second ed.

\bibitem{Pechukas81}
\bibinfo{author}{\bibfnamefont{P.}~\bibnamefont{Pechukas}},
  \bibinfo{journal}{Ann. Rev. Phys. Chem.} \textbf{\bibinfo{volume}{32}},
  \bibinfo{pages}{159} (\bibinfo{year}{1981}).

\bibitem{Langevin1905}
\bibinfo{author}{\bibfnamefont{P.}~\bibnamefont{Langevin}},
  \bibinfo{journal}{Annal. Chim. Phys.} \textbf{\bibinfo{volume}{5}},
  \bibinfo{pages}{245} (\bibinfo{year}{1905}).

\bibitem{Chesnavich82}
\bibinfo{author}{\bibfnamefont{W.~J.} \bibnamefont{Chesnavich}}
  \bibnamefont{and} \bibinfo{author}{\bibfnamefont{M.~T.}
  \bibnamefont{Bowers}}, \emph{\bibinfo{title}{{Theory of Ion-Neutral
  Interactions: Application of Transition State Theory Concepts to Both
  Collisional and Reactive Properties of Simple Systems}}}
  (\bibinfo{publisher}{Pergamon Press}, \bibinfo{year}{1982}).

\bibitem{Miller76a}
\bibinfo{author}{\bibfnamefont{W.~H.} \bibnamefont{Miller}},
  \bibinfo{journal}{J. Chem. Phys.} \textbf{\bibinfo{volume}{65}},
  \bibinfo{pages}{2216} (\bibinfo{year}{1976}).

\bibitem{Chesnavich1986}
\bibinfo{author}{\bibfnamefont{W.~J.} \bibnamefont{Chesnavich}},
  \bibinfo{journal}{J. Chem. Phys.} \textbf{\bibinfo{volume}{84}},
  \bibinfo{pages}{2615} (\bibinfo{year}{1986}).

\bibitem{MauguiereCPL2014}
\bibinfo{author}{\bibfnamefont{F.~A.~L.} \bibnamefont{Mauguiere}},
  \bibinfo{author}{\bibfnamefont{P.}~\bibnamefont{Collins}},
  \bibinfo{author}{\bibfnamefont{G.~S.} \bibnamefont{Ezra}},
  \bibinfo{author}{\bibfnamefont{S.~C.} \bibnamefont{Farantos}},
  \bibnamefont{and} \bibinfo{author}{\bibfnamefont{S.}~\bibnamefont{Wiggins}},
  \bibinfo{journal}{Chem. Phys. Lett.} \textbf{\bibinfo{volume}{592}}, \bibinfo{pages}{282-27}
  (\bibinfo{year}{2014}).

\bibitem{Wiesenfeld05}
\bibinfo{author}{\bibfnamefont{L.}~\bibnamefont{Wiesenfeld}},
  \bibinfo{journal}{Adv. Chem. Phys.} \textbf{\bibinfo{volume}{130 A}},
  \bibinfo{pages}{217} (\bibinfo{year}{2005}).

\bibitem{Thiele62}
\bibinfo{author}{\bibfnamefont{E.}~\bibnamefont{Thiele}}, \bibinfo{journal}{J.
  Chem. Phys.} \textbf{\bibinfo{volume}{36}}(\bibinfo{number}{6}),
  \bibinfo{pages}{1466} (\bibinfo{year}{1962}).

\bibitem{Thiele62a}
\bibinfo{author}{\bibfnamefont{E.}~\bibnamefont{Thiele}}, \bibinfo{journal}{J.
  Chem. Phys.} \textbf{\bibinfo{volume}{38}}(\bibinfo{number}{8}),
  \bibinfo{pages}{1959} (\bibinfo{year}{1963}).

\bibitem{VandeLinde90a}
\bibinfo{author}{\bibfnamefont{S.~R.} \bibnamefont{{Vande Linde}}}
  \bibnamefont{and} \bibinfo{author}{\bibfnamefont{W.~L.} \bibnamefont{Hase}},
  \bibinfo{journal}{J. Phys. Chem.}
  \textbf{\bibinfo{volume}{94}}(\bibinfo{number}{16}), \bibinfo{pages}{6148}
  (\bibinfo{year}{1990}).

\bibitem{Peslherbe95}
\bibinfo{author}{\bibfnamefont{G.~H.} \bibnamefont{Peslherbe}},
  \bibinfo{author}{\bibfnamefont{H.}~\bibnamefont{Wang}}, \bibnamefont{and}
  \bibinfo{author}{\bibfnamefont{W.~L.} \bibnamefont{Hase}},
  \bibinfo{journal}{J. Chem. Phys.}
  \textbf{\bibinfo{volume}{102}}(\bibinfo{number}{14}), \bibinfo{pages}{5626}
  (\bibinfo{year}{1995}),
  \urlprefix\url{http://scitation.aip.org/content/aip/journal/jcp/102/14/10.10%
63/1.469294}.

\bibitem{Klippenstein2011}
\bibinfo{author}{\bibfnamefont{S.~J.} \bibnamefont{Klippenstein}},
  \bibinfo{author}{\bibfnamefont{Y.}~\bibnamefont{Georgievskii}},
  \bibnamefont{and} \bibinfo{author}{\bibfnamefont{L.~B.}
  \bibnamefont{Harding}}, \bibinfo{journal}{J. Phys. Chem. A}
  \textbf{\bibinfo{volume}{115}}, \bibinfo{pages}{14370}
  (\bibinfo{year}{2011}).

\bibitem{Chesnavich1981}
\bibinfo{author}{\bibfnamefont{W.~J.} \bibnamefont{Chesnavich}},
  \bibinfo{author}{\bibfnamefont{L.}~\bibnamefont{Bass}},
  \bibinfo{author}{\bibfnamefont{T.}~\bibnamefont{Su}}, \bibnamefont{and}
  \bibinfo{author}{\bibfnamefont{M.~T.} \bibnamefont{Bowers}},
  \bibinfo{journal}{J. Chem. Phys.}
  \textbf{\bibinfo{volume}{74}}(\bibinfo{number}{4}), \bibinfo{pages}{2228}
  (\bibinfo{year}{1981}).

\bibitem{Pechukas73}
\bibinfo{author}{\bibfnamefont{P.}~\bibnamefont{Pechukas}} \bibnamefont{and}
  \bibinfo{author}{\bibfnamefont{F.~J.} \bibnamefont{McLafferty}},
  \bibinfo{journal}{J. Chem. Phys.}
  \textbf{\bibinfo{volume}{58}}(\bibinfo{number}{4}), \bibinfo{pages}{1622}
  (\bibinfo{year}{1973}),
  \urlprefix\url{http://scitation.aip.org/content/aip/journal/jcp/58/4/10.1063%
/1.1679404}.

\bibitem{Pechukas77}
\bibinfo{author}{\bibfnamefont{P.}~\bibnamefont{Pechukas}} \bibnamefont{and}
  \bibinfo{author}{\bibfnamefont{E.}~\bibnamefont{Pollak}},
  \bibinfo{journal}{J. Chem. Phys.}
  \textbf{\bibinfo{volume}{67}}(\bibinfo{number}{12}), \bibinfo{pages}{5976}
  (\bibinfo{year}{1977}),
  \urlprefix\url{http://scitation.aip.org/content/aip/journal/jcp/67/12/10.106%
3/1.434777}.

\bibitem{Pollak78}
\bibinfo{author}{\bibfnamefont{E.}~\bibnamefont{Pollak}} \bibnamefont{and}
  \bibinfo{author}{\bibfnamefont{P.}~\bibnamefont{Pechukas}},
  \bibinfo{journal}{J. Chem. Phys.} \textbf{\bibinfo{volume}{69}},
  \bibinfo{pages}{1218} (\bibinfo{year}{1978}).

\bibitem{Pechukas79}
\bibinfo{author}{\bibfnamefont{P.}~\bibnamefont{Pechukas}} \bibnamefont{and}
  \bibinfo{author}{\bibfnamefont{E.}~\bibnamefont{Pollak}},
  \bibinfo{journal}{J. Chem. Phys.}
  \textbf{\bibinfo{volume}{71}}(\bibinfo{number}{5}), \bibinfo{pages}{2062}
  (\bibinfo{year}{1979}),
  \urlprefix\url{http://scitation.aip.org/content/aip/journal/jcp/71/5/10.1063%
/1.438575}.

\bibitem{waalkens2004direct}
\bibinfo{author}{\bibfnamefont{H.}~\bibnamefont{Waalkens}} \bibnamefont{and}
  \bibinfo{author}{\bibfnamefont{S.}~\bibnamefont{Wiggins}},
  \bibinfo{journal}{J. Phys. A: Math. Gen.}
  \textbf{\bibinfo{volume}{37}}(\bibinfo{number}{35}), \bibinfo{pages}{L435}
  (\bibinfo{year}{2004}).

\bibitem{wiggins2001impenetrable}
\bibinfo{author}{\bibfnamefont{S.}~\bibnamefont{Wiggins}},
  \bibinfo{author}{\bibfnamefont{L.}~\bibnamefont{Wiesenfeld}},
  \bibinfo{author}{\bibfnamefont{C.}~\bibnamefont{Jaff{\'e}}},
  \bibinfo{author}{\bibfnamefont{T.}~\bibnamefont{Uzer}}, \emph{et~al.},
  \bibinfo{journal}{Phys. Rev. Lett.}
  \textbf{\bibinfo{volume}{86}}(\bibinfo{number}{24}), \bibinfo{pages}{5478}
  (\bibinfo{year}{2001}).

\bibitem{Fenichel1971}
\bibinfo{author}{\bibfnamefont{N.}~\bibnamefont{Fenichel}},
  \bibinfo{journal}{Indiana Univ. Math. J}
  \textbf{\bibinfo{volume}{21}}(\bibinfo{number}{193-226}),
  \bibinfo{pages}{1972} (\bibinfo{year}{1971}).

\bibitem{Fenichel1977}
\bibinfo{author}{\bibfnamefont{N.}~\bibnamefont{Fenichel}},
  \bibinfo{journal}{Indiana Univ. Math. J.} \textbf{\bibinfo{volume}{26}},
  \bibinfo{pages}{81} (\bibinfo{year}{1977}), ISSN \bibinfo{issn}{0022-2518}.

\bibitem{Fenichel1974}
\bibinfo{author}{\bibfnamefont{N.}~\bibnamefont{Fenichel}},
  \bibinfo{journal}{Indiana Univ. Math. J.} \textbf{\bibinfo{volume}{23}},
  \bibinfo{pages}{1109} (\bibinfo{year}{1974}), ISSN \bibinfo{issn}{0022-2518}.

\bibitem{Farantos09}
\bibinfo{author}{\bibfnamefont{S.~C.} \bibnamefont{Farantos}},
  \bibinfo{author}{\bibfnamefont{R.}~\bibnamefont{Schinke}},
  \bibinfo{author}{\bibfnamefont{H.}~\bibnamefont{Guo}}, \bibnamefont{and}
  \bibinfo{author}{\bibfnamefont{M.}~\bibnamefont{Joyeux}},
  \bibinfo{journal}{Chem. Rev.}
  \textbf{\bibinfo{volume}{109}}(\bibinfo{number}{9}), \bibinfo{pages}{4248}
  (\bibinfo{year}{2009}).

\bibitem{Mauguiere10}
\bibinfo{author}{\bibfnamefont{F.}~\bibnamefont{Mauguiere}},
  \bibinfo{author}{\bibfnamefont{M.}~\bibnamefont{Rey}},
  \bibinfo{author}{\bibfnamefont{V.}~\bibnamefont{Tyuterev}},
  \bibinfo{author}{\bibfnamefont{J.}~\bibnamefont{Suarez}}, \bibnamefont{and}
  \bibinfo{author}{\bibfnamefont{S.~C.} \bibnamefont{Farantos}},
  \bibinfo{journal}{J. Phys. Chem. A}
  \textbf{\bibinfo{volume}{114}}(\bibinfo{number}{36}), \bibinfo{pages}{9836}
  (\bibinfo{year}{2010}).

\bibitem{Farantos98}
\bibinfo{author}{\bibfnamefont{S.~C.} \bibnamefont{Farantos}},
  \bibinfo{journal}{Comput. Phys. Commun.} \textbf{\bibinfo{volume}{108}},
  \bibinfo{pages}{240} (\bibinfo{year}{1998}).

\bibitem{pw94}
\bibinfo{author}{\bibfnamefont{A.~D.} \bibnamefont{Perry}} \bibnamefont{and}
  \bibinfo{author}{\bibfnamefont{S.}~\bibnamefont{Wiggins}},
  \bibinfo{journal}{Physica D}
  \textbf{\bibinfo{volume}{71}}(\bibinfo{number}{1-2}), \bibinfo{pages}{102}
  (\bibinfo{year}{1994}).

\bibitem{mg95b}
\bibinfo{author}{\bibfnamefont{A.}~\bibnamefont{Morbidelli}} \bibnamefont{and}
  \bibinfo{author}{\bibfnamefont{A.}~\bibnamefont{Giorgilli}},
  \bibinfo{journal}{J. Stat. Phys.}
  \textbf{\bibinfo{volume}{78}}(\bibinfo{number}{5-6}), \bibinfo{pages}{1607}
  (\bibinfo{year}{1995}).

\bibitem{Harding07}
\bibinfo{author}{\bibfnamefont{L.~B.} \bibnamefont{Harding}},
  \bibinfo{author}{\bibfnamefont{S.~J.} \bibnamefont{Klippenstein}},
  \bibnamefont{and} \bibinfo{author}{\bibfnamefont{A.~W.}
  \bibnamefont{Jasper}}, \bibinfo{journal}{Phys. Chem. Chem. Phys.}
  \textbf{\bibinfo{volume}{9}}, \bibinfo{pages}{4055} (\bibinfo{year}{2007}).

\bibitem{Harding10}
\bibinfo{author}{\bibfnamefont{L.~B.} \bibnamefont{Harding}} \bibnamefont{and}
  \bibinfo{author}{\bibfnamefont{S.~J.} \bibnamefont{Klippenstein}},
  \bibinfo{journal}{J. Phys. Chem. Lett.}
  \textbf{\bibinfo{volume}{1}}(\bibinfo{number}{20}), \bibinfo{pages}{3016}
  (\bibinfo{year}{2010}).

\bibitem{Newhouse79}
\bibinfo{author}{\bibfnamefont{S.~E.} \bibnamefont{Newhouse}},
  \bibinfo{journal}{Publ. Math. IHES} \textbf{\bibinfo{volume}{50}},
  \bibinfo{pages}{101} (\bibinfo{year}{1979}).

\bibitem{Duarte99}
\bibinfo{author}{\bibfnamefont{P.}~\bibnamefont{Duarte}},
  \bibinfo{journal}{Dyn. Stab. Sys.}
  \textbf{\bibinfo{volume}{14}}(\bibinfo{number}{4}), \bibinfo{pages}{339}
  (\bibinfo{year}{1999}).

\bibitem{Gonchenko00}
\bibinfo{author}{\bibfnamefont{S.~V.} \bibnamefont{Gonchenko}}
  \bibnamefont{and} \bibinfo{author}{\bibfnamefont{L.~P.}
  \bibnamefont{Silnikov}}, \bibinfo{journal}{J. Stat. Phys.}
  \textbf{\bibinfo{volume}{101}}(\bibinfo{number}{1/2}), \bibinfo{pages}{321}
  (\bibinfo{year}{2000}).

\bibitem{Farantos11}
\bibinfo{author}{\bibfnamefont{F.~A.~L.} \bibnamefont{Mauguiere}},
  \bibinfo{author}{\bibfnamefont{S.~C.} \bibnamefont{Farantos}},
  \bibinfo{author}{\bibfnamefont{J.}~\bibnamefont{Suarez}}, \bibnamefont{and}
  \bibinfo{author}{\bibfnamefont{R.}~\bibnamefont{Schinke}},
  \bibinfo{journal}{J. Chem. Phys.}
  \textbf{\bibinfo{volume}{134}}(\bibinfo{number}{24}), \bibinfo{pages}{244302}
  (\bibinfo{year}{2011}).

\bibitem{Slater56}
\bibinfo{author}{\bibfnamefont{N.~B.} \bibnamefont{Slater}},
  \bibinfo{journal}{J. Chem. Phys.}
  \textbf{\bibinfo{volume}{24}}(\bibinfo{number}{6}), \bibinfo{pages}{1256}
  (\bibinfo{year}{1956}).

\bibitem{Slater59}
\bibinfo{author}{\bibfnamefont{N.~B.} \bibnamefont{Slater}},
  \emph{\bibinfo{title}{{Theory of Unimolecular Reactions}}}
  (\bibinfo{publisher}{Cornell University Press}, \bibinfo{address}{Ithaca,
  NY}, \bibinfo{year}{1959}).

\bibitem{Bunker62}
\bibinfo{author}{\bibfnamefont{D.~L.} \bibnamefont{Bunker}},
  \bibinfo{journal}{J. Chem. Phys.} \textbf{\bibinfo{volume}{37}},
  \bibinfo{pages}{393} (\bibinfo{year}{1962}).

\bibitem{Bunker64}
\bibinfo{author}{\bibfnamefont{D.~L.} \bibnamefont{Bunker}},
  \bibinfo{journal}{J. Chem. Phys.} \textbf{\bibinfo{volume}{40}},
  \bibinfo{pages}{1946} (\bibinfo{year}{1964}).

\bibitem{Bunker73}
\bibinfo{author}{\bibfnamefont{D.~L.} \bibnamefont{Bunker}} \bibnamefont{and}
  \bibinfo{author}{\bibfnamefont{W.~L.} \bibnamefont{Hase}},
  \bibinfo{journal}{J. Chem. Phys.} \textbf{\bibinfo{volume}{59}},
  \bibinfo{pages}{4621} (\bibinfo{year}{1973}).

\bibitem{Dumont86}
\bibinfo{author}{\bibfnamefont{R.~S.} \bibnamefont{Dumont}} \bibnamefont{and}
  \bibinfo{author}{\bibfnamefont{P.}~\bibnamefont{Brumer}},
  \bibinfo{journal}{J. Phys. Chem.} \textbf{\bibinfo{volume}{90}},
  \bibinfo{pages}{3509} (\bibinfo{year}{1986}).

\bibitem{DeLeon81}
\bibinfo{author}{\bibfnamefont{N.}~\bibnamefont{DeLeon}} \bibnamefont{and}
  \bibinfo{author}{\bibfnamefont{B.~J.} \bibnamefont{Berne}},
  \bibinfo{journal}{J. Chem. Phys.} \textbf{\bibinfo{volume}{75}},
  \bibinfo{pages}{3495} (\bibinfo{year}{1981}).

\bibitem{Berne82}
\bibinfo{author}{\bibfnamefont{B.~J.} \bibnamefont{Berne}},
  \bibinfo{author}{\bibfnamefont{N.}~\bibnamefont{DeLeon}}, \bibnamefont{and}
  \bibinfo{author}{\bibfnamefont{R.~O.} \bibnamefont{Rosenberg}},
  \bibinfo{journal}{J. Phys. Chem.} \textbf{\bibinfo{volume}{86}},
  \bibinfo{pages}{2166} (\bibinfo{year}{1982}).

\bibitem{Binney85}
\bibinfo{author}{\bibfnamefont{J.}~\bibnamefont{Binney}},
  \bibinfo{author}{\bibfnamefont{O.~E.} \bibnamefont{Gerhard}},
  \bibnamefont{and} \bibinfo{author}{\bibfnamefont{P.}~\bibnamefont{Hut}},
  \bibinfo{journal}{Mon. Not. Roy. Astron. Soc.}
  \textbf{\bibinfo{volume}{215}}, \bibinfo{pages}{59} (\bibinfo{year}{1985}).

\bibitem{Meyer86}
\bibinfo{author}{\bibfnamefont{H.-D.} \bibnamefont{Meyer}},
  \bibinfo{journal}{J. Chem. Phys.} \textbf{\bibinfo{volume}{84}},
  \bibinfo{pages}{3147} (\bibinfo{year}{1986}).

\bibitem{Brumer80}
\bibinfo{author}{\bibfnamefont{P.}~\bibnamefont{Brumer}},
  \bibinfo{author}{\bibfnamefont{D.~E.} \bibnamefont{Fitz}}, \bibnamefont{and}
  \bibinfo{author}{\bibfnamefont{D.}~\bibnamefont{Wardlaw}},
  \bibinfo{journal}{J. Chem. Phys.}
  \textbf{\bibinfo{volume}{72}}(\bibinfo{number}{1}), \bibinfo{pages}{386}
  (\bibinfo{year}{1980}).

\bibitem{Pollak81}
\bibinfo{author}{\bibfnamefont{E.}~\bibnamefont{Pollak}}, \bibinfo{journal}{J.
  Chem. Phys.} \textbf{\bibinfo{volume}{74}}, \bibinfo{pages}{6763}
  (\bibinfo{year}{1981}).

\bibitem{Waalkens05}
\bibinfo{author}{\bibfnamefont{H.}~\bibnamefont{Waalkens}},
  \bibinfo{author}{\bibfnamefont{A.}~\bibnamefont{Burbanks}}, \bibnamefont{and}
  \bibinfo{author}{\bibfnamefont{S.}~\bibnamefont{Wiggins}},
  \bibinfo{journal}{Phys. Rev. Lett.} \textbf{\bibinfo{volume}{95}},
  \bibinfo{pages}{Art. No. 084301} (\bibinfo{year}{2005}).

\bibitem{Waalkens05a}
\bibinfo{author}{\bibfnamefont{H.}~\bibnamefont{Waalkens}},
  \bibinfo{author}{\bibfnamefont{A.}~\bibnamefont{Burbanks}}, \bibnamefont{and}
  \bibinfo{author}{\bibfnamefont{S.}~\bibnamefont{Wiggins}},
  \bibinfo{journal}{J. Phys. A} \textbf{\bibinfo{volume}{38}},
  \bibinfo{pages}{L759} (\bibinfo{year}{2005}).

\bibitem{Hase83}
\bibinfo{author}{\bibfnamefont{W.~L.} \bibnamefont{Hase}},
  \bibinfo{author}{\bibfnamefont{D.~G.} \bibnamefont{Buckowski}},
  \bibnamefont{and} \bibinfo{author}{\bibfnamefont{K.~N.} \bibnamefont{Swamy}},
  \bibinfo{journal}{J. Phys. Chem.} \textbf{\bibinfo{volume}{87}},
  \bibinfo{pages}{2754} (\bibinfo{year}{1983}).

\bibitem{Grebenshchikov03}
\bibinfo{author}{\bibfnamefont{S.~Y.} \bibnamefont{Grebenshchikov}},
  \bibinfo{author}{\bibfnamefont{R.}~\bibnamefont{Schinke}}, \bibnamefont{and}
  \bibinfo{author}{\bibfnamefont{W.~L.} \bibnamefont{Hase}}, in
  \emph{\bibinfo{booktitle}{{Unimolecular Kinetics: Part 1. The Reaction
  Step}}}, edited by \bibinfo{editor}{\bibfnamefont{N.~J.~B.}
  \bibnamefont{Greene}} (\bibinfo{publisher}{Elsevier}, \bibinfo{address}{New
  York}, \bibinfo{year}{2003}), vol.~\bibinfo{volume}{{39}} of
  \emph{\bibinfo{series}{{Comprehensive Chemical Kinetics}}}, pp.
  \bibinfo{pages}{105--242}.

\bibitem{Berblinger94}
\bibinfo{author}{\bibfnamefont{M.}~\bibnamefont{Berblinger}} \bibnamefont{and}
  \bibinfo{author}{\bibfnamefont{C.}~\bibnamefont{Schlier}},
  \bibinfo{journal}{J. Chem. Phys.} \textbf{\bibinfo{volume}{101}},
  \bibinfo{pages}{4750} (\bibinfo{year}{1994}).

\bibitem{Gray87}
\bibinfo{author}{\bibfnamefont{S.~K.} \bibnamefont{Gray}} \bibnamefont{and}
  \bibinfo{author}{\bibfnamefont{S.~A.} \bibnamefont{Rice}},
  \bibinfo{journal}{J. Chem. Phys.} \textbf{\bibinfo{volume}{86}},
  \bibinfo{pages}{2020} (\bibinfo{year}{1987}).

\bibitem{Stember07}
\bibinfo{author}{\bibfnamefont{J.~N.} \bibnamefont{Stember}} \bibnamefont{and}
  \bibinfo{author}{\bibfnamefont{G.~S.} \bibnamefont{Ezra}},
  \bibinfo{journal}{Chem. Phys.} \textbf{\bibinfo{volume}{337}},
  \bibinfo{pages}{11} (\bibinfo{year}{2007}).

\bibitem{Nagahata13}
\bibinfo{author}{\bibfnamefont{Y.}~\bibnamefont{Nagahata}},
  \bibinfo{author}{\bibfnamefont{H.}~\bibnamefont{Teramoto}},
  \bibinfo{author}{\bibfnamefont{C.}~\bibnamefont{Li}},
  \bibinfo{author}{\bibfnamefont{S.}~\bibnamefont{Kawai}}, \bibnamefont{and}
  \bibinfo{author}{\bibfnamefont{T.}~\bibnamefont{Komatsuzaki}},
  \bibinfo{journal}{Phys. Rev. E} \textbf{\bibinfo{volume}{88}},
  \bibinfo{pages}{042923} (\bibinfo{year}{2013}).

\bibitem{Grice87}
\bibinfo{author}{\bibfnamefont{M.}~\bibnamefont{Grice}},
  \bibinfo{author}{\bibfnamefont{B.}~\bibnamefont{Andrews}}, \bibnamefont{and}
  \bibinfo{author}{\bibfnamefont{W.}~\bibnamefont{Chesnavich}},
  \bibinfo{journal}{J. Chem. Phys.} \textbf{\bibinfo{volume}{87}},
  \bibinfo{pages}{959} (\bibinfo{year}{1987}).

\bibitem{Mauguiere13}
\bibinfo{author}{\bibfnamefont{F.}~\bibnamefont{Mauguiere}},
  \bibinfo{author}{\bibfnamefont{P.}~\bibnamefont{Collins}},
  \bibinfo{author}{\bibfnamefont{G.}~\bibnamefont{Ezra}}, \bibnamefont{and}
  \bibinfo{author}{\bibfnamefont{S.}~\bibnamefont{Wiggins}},
  \bibinfo{journal}{J. Chem. Phys.} \textbf{\bibinfo{volume}{138}},
  \bibinfo{pages}{134118} (\bibinfo{year}{2013}).

\end{thebibliography}

\newpage

\newpage

\newpage

\newpage

\newpage

\newpage

\newpage

\newpage

\newpage

\newpage

\newpage

\newpage

\newpage

\newpage

\end{document}